\documentclass[12pt]{article}
\usepackage{amsmath,amssymb}
 \usepackage{graphicx}
\begin{document}

\begin{center}

{\Large \bf Spheroidal and torsional modes of quasistatic shear
oscillations in
 the solid globe models of nuclear physics and pulsar
astrophysics}

\vspace{0.7cm}

 {S. Bastrukov$^1$, H-K. Chang$^2$, \c S. Mi\c sicu$^3$, I. Molodtsova$^1$, D. Podgainy$^1$}

\vspace{0.5cm}

\noindent {\it  $^1$ Laboratory of Informational Technologies,
Joint Institute for
Nuclear Research, \\
141980 Dubna, Moscow Region, Russia\\

\noindent $^2$ Department of Physics and  Institute of Astronomy,
  National Tsing Hua University, Hsinchu, 30013, Taiwan, Republic of China\\

\noindent $^3$ Department of Theoretical Physics, National
Institute for Nuclear Research,
 Buchurest, P.O. Box MG6, Romania

}

\end{center}

\begin{abstract}
 The past three decades of investigation on nuclear physics and pulsar astrophysics have seen gradual recognition
 that elastodynamic approach to the continuum mechanics of nuclear matter
 provides proper account of macroscopic motions of degenerate Fermi-matter constituting interior of
 the nuclear material objects, the densest of all known today.
 This paper focuses on one theoretical issue of this development which is concerned with oscillatory behavior
 of a viscoelastic solid globe in the regime of quasistatic, force-free, non-compressional oscillations less
 investigated in the literature compared to oscillations in the regime of standing shear waves.
 We show that in this case the problem of computing frequency and lifetime of spheroidal
 and torsional modes of non-radial shear vibrations damped by viscosity can be unambiguously resolved by working
 from the energy balance equation and taking advantage of the Rayleigh's variational method. The efficiency of this
 method is demonstrated by solid globe models of nuclear physics and pulsar astrophysics
 dealing with oscillations of a spherical mass of a viscoelastic Fermi-solid with homogeneous and non-homogeneous
 profiles of the bulk density, the shear modulus, and the shear viscosity.
\end{abstract}


\noindent


\vspace{0.2cm}

\noindent {\sl Preprint of an article accepted for publication in
International Journal of Modern Physics A: Particles and Fields;
Gravitation; Cosmology; Nuclear Physics [2007]}

\newpage

\section{Introduction}
 An understanding dynamical laws governing macroscopic motions
 of degenerate nucleonic material -- the continuum mechanics of nuclear matter
 --
 is important for developing interconnected view of the nuclear physics and pulsar
 astrophysics. The nearly linear
 dependence of the nucleus volume and the nucleus binding energy upon number of
 nucleons, meaning the saturation of internucleon forces, unambiguously indicate that this "most conspicuous feature
 of nuclei" (Bethe 1999) is the same as it is for condensed matter. The condensed
 matter, as is known, comes in two forms -- liquid and solid. Accordingly, the question which of two fundamental models of
 material continua -- the fluid-mechanical (relying on equations of
 hydrodynamics highlighting fluidity as the major dynamical property of liquid state of condensed matter) or the
 solid-mechanical
 (resting on equations of elastodynamics accentuating reversal elastic deformations as basic dynamical property of
 solid state of condensed matter) can provide adequate description
 of the observable behavior of the nuclear matter objects has been and still is central to
 the program on the study of material properties of superdense substance constituting interior of atomic nuclei
 and neutron stars.

 It is of common knowledge today that the earlier liquid drop model of nucleus, implying the fluid-mechanical pattern
 of continuous nuclear matter and complete disorder in the behavior of constituting particles,
 has fallen in disfavour after discovery of structural order in proton-neutron composition of nucleus which received
 theoretical interpretation in the nuclear shell model. This microscopic signature of
 ordered clusterization of nucleons together with such macroscopic one as the deformed equilibrium
 shapes indicate that the atomic nucleus bears more resemblance to an ultrafine particle of a
 viscoelastic solid than a drop of liquid which is characterized by spherical equilibrium shape.
 Such an understanding is clearly traced in the macroscopic approaches
 to the  nuclear collective dynamics and asteroseismology of neutron stars developed in the past three decades
 that have been dominated by mathematical methods and physical notions of the classical elastodynamics testifying
 gradual receding of the fluid-mechanical paradigm of the nuclear continuum mechanics into the solid mechanical
 one. While the microscopic nature of structural order underlying substantially
 quantum clusterization of nucleons in the nuclear matter objects, which has definitely nothing to do
 with classical picture of lattice-like crystallization of atoms in ordinary solids, is still to be clarified\footnote{In
 current investigations on quantum condensed media,
 the term Fermi-solid (which is, as far as we  know, due to Guyer
 (1973)) is applied to accentuate the difference between single-particle behavior of
 microconstituents in quantum and ordinary solids (e.g. Polturak and Gov
 2003).}, an independent inference of current nuclear physics
 and pulsar astrophysics is that the elastodynamical model of nuclear continuum
 mechanics, lying at the base of the solid globe model for an atomic nucleus and the solid
 star model for a pulsar, provides proper macroscopic account of data on giant-resonant oscillatory
 response of nuclei and non-radial pulsations of neutron stars.

 By now there are quite cogent theoretical arguments showing that the asteroseismology of neutron stars, studying
 restless behavior of pulsars and magnetars, can be properly understood working from the model of a
 solid star\footnote{To the best of our knowledge the term
 "Solid Stars" for neutron stars has been suggested by M. Ruderman  in the infancy of pulsar astrophysics,
 as pointed out in (Harwit 2006). A current view on the neutron star structure is extensively discussed in
 (Shapiro and Teukolsky 1983; Weber 1999; Glendenning 2000; Lattimer and Prakash 2001).}, rather
 than the liquid star model underlying current understanding of seismic activity of the main sequence
 stars. Also, it is believed the much more denser strange matter of quark stars (e.g. Weber 2005) is
 in the solid state too (Xu 2005; Owen 2005).
 In broad terms, by a solid star is understood a self-gravitating mass of viscoelastic solid
 continuum, highly robust to compressional deformations,  whose motions are governed by the laws of Newtonian
 elastodynamics. In pulsar astrophysics, a model of a solid
 star undergoing global and surface (crustal) elastic vibrations
 has been invoked to explain microspikes detected in
 the windows of the rotation driven main pulse train of the pulsar
 radio emission
 (e.g.  Van Horn 1980, McDermott, Van Horn and Hansen 1988; Strohmayer, Cordes and van Horn 1993; Bastrukov, Weber and
 Podgainy 1999). The core-crust model of a solid star was found to provide proper
 account of the neutron star quakes manifested by glitches in the regular pulsating emission of radio pulsars
 (e.g. Pines, Shaham and Ruderman 1974; Canuto and Chitre
 1974; Franco, Link and Epstein 2000) and gamma bursts of pulsars and magnetars
 (Blaes, Blandford, Goldreich, Madau 1989; Cheng, Epstein, Guyer and Young 1996). In this model the neutron
 star interior is thought of as a stiff outer crust covering a denser core which are separated by thin inner
 crust whose matter operates like a lubricant facilitating tangential shear displacements of the outer crust against core.
 This picture suggests that the post-quake recovering of glitching radio pulsars and gamma-bursting magnetars
 to the regime
 of quiescent pulsed radiation is accompanied by the torsional, that is, differentially-rotational elastic vibrations
 of the neutron star crust against core. Over the years the torsional pulsations of neutron stars have gained
 considerable attention in different aspects of radiative activity of pulsars and magnetars
 (e.g. Ruderman 1969; Van Horn 1980, Hansen and Chioffi 1980; Bastrukov, Podgainy,
 Yang, Weber 2002). This kind of vibrations, generic to
 oscillatory behavior of an elastic solid globe, is one of the topical issues in current discussion
 of periodicity in bursting activity of
 X-ray pulsars (e.g. Lin and Chang 2005, Strohmayer and Watts 2006;
 Strohmayer and Bildsten 2006; Bisnovatyi-Kogan 2007).

 In nuclear physics, the signatures typical of oscillatory behavior
 of viscoelastic solid globe have been discovered in the high-energy collective response
 of atomic nucleus by giant resonances (Bertsch 1974) -- the fundamental modes of vibrational
 excitations generic to all stable nuclei of the periodic table (e.g. Harakeh and van der
 Woude 2003; Richter 2005). Theoretical treatment of these modes in terms of spheroidal (electric resonances)
 and torsional (magnetic resonances) elastic vibrations of a solid
 sphere has been and still is the subject of extensive research by
 different methods of the nuclear collective dynamics
 (e.g. Semenko 1977; Sagawa and Holzwarth 1978; Nix and Sierk 1980; Wong and Azziz 1981;
 Hasse, Ghosh, Winter and Lumbroso 1982; Stringari 1983; Balbutsev and Mikhailov
 1988; Kolomietz 1990; Di Toro 1991; Bastrukov,  Mi\c sicu and Sushkov 1993;  S. Mi\c sicu 2006 and references therein).
 One of the main goals of these investigations  is to get insight into material laws, that is the rheology of the
 nuclear matter. The torsional oscillatory mode is unique to the nuclear
 solid globe model relying on equations of the solid mechanics. Thereby, the macroscopic
 interpretation of the giant magnetic resonances discovered in the
 photoelectron experiments (Raman, Fagg and Hicks, 1991;
 Luttge, von Neumann-Cosel, Neumeyer and Richter 1996) as manifestation of the nuclear
 twist (Holzwarth and Eckart 1977; Bastrukov and Gudkov 1992), that is, torsional differentially-rotational elastic
 oscillations of a nucleus (Bastrukov, Molodtsova and Shilov 1995; Bastrukov, Libert and  Molodtsova
 1997) serves as the most strong argument in favor of the elastodynamical model of nuclear continuum
 mechanics. The equations of fluid mechanics
 underlying canonical liquid drop model of nucleus (e.g. Greiner and Maruhn 1996) have no solutions
 describing nuclear response by differentially rotational oscillations
 and this is the reason why the liquid drop model excludes vibrational states of magnetic type, as was pointed
 out long ago in (Rowe 1970).

 This paper focuses on one subtle issue of general solid-mechanical treatment
 of oscillatory behavior of the nuclear matter objects in terms of vibrations of a viscoelastic solid globe
 that has been called into question by the above development of both the nuclear physics and pulsar astrophysics.
 Specifically, it is concerned with the regime of quasistatic, force-free, oscillations less investigated
 in the literature as compared to the solid globe oscillations in the regime of standing
 shear waves. The most striking feature of the quasistatic regime is that the restoring force of
 Hookean elastic (reversal) stress and dissipative force of Newtonian viscous (irreversal) stress entering
 the basic equation of solid mechanics turn to zero (from what the term quasistatic oscillations is
 derived), but the material stresses themselves and the work done by these stresses in the bulk of an oscillating solid
 globe do not. Based on this observation we show that in the case of force-free fluctuations of
 stresses the frequency and lifetime of both spheroidal and torsional modes
 in a viscoelastic solid globe can be computed by taking advantage of the energy variational principle relying
 on the equation of energy conservation, not on the equation of
 elastodynamics as is the case of oscillations in the regime of standing waves.

 The paper is organized as follows. In Section 2 a brief outline
 is given of solid-mechanical equations modeling
 oscillatory behavior of a viscoelastic solid with focus on the problem of oscillatory response
 of a perfectly elastic (non-viscous) sphere
 in the regime of standing shear waves that dates back to the canonical work of Lamb (1882).
 Section 3 is devoted to the solution of the problem of quasistatic
 oscillations of a viscoelastic solid globe by  Rayleigh's variational method and its
 application to the nuclear giant resonances.
 The efficiency of this method is demonstrated in Section  4  considering several mathematical models of astrophysical
 interest. The practical usefulness of these models is briefly discussed in
 Summary and details of calculations are presented in Appendices.

\section{Elastodynamics of the solid globe oscillations in the compression free regime of standing shear waves}

 In approaching the topic like this, it seems worthy to start with a brief outline of physical ideas
 and mathematical equations underlying the standard continuum model of
 solid mechanics or elastodynamics (e.g. Love 1944; Morse and Feshbach 1953; Eringen and Suhubi 1975; Landau,
 Lifshits, Kosevich and Pitaevskii 1995).
 The basic dynamical variable characterizing the state of motion of solid continuous medium is the field
 of material displacement $u_i({\bf r},t)$ whose  emergence in the volume of solid object
 is associated with its response to perturbation of mechanically equilibrium state in which $u_i=0$.
 The canonical continuum model of an isotropic viscoelastic solid rests on two fundamental,
 empirically verified, statements. First is the generalized Hookean law of elastic deformations which is
 expressed by linear relation between
 the symmetric tensors of elastic (reversal) stress $p_{ik}$ and shear strain or deformation
 $u_{ik}$:
  \begin{eqnarray}
 \label{e2.1}
 && p_{ik}=2\mu u_{ik}+
 \left(\kappa-\frac{2}{3}\mu\right)\,u_{jj}\delta_{ik}=2\mu \left(u_{ik}-\frac{1}{3}u_{jj}\delta_{ik}\right)+
 \kappa\,u_{jj}\delta_{ik}\\
 \label{e2.2}
 && u_{ik}=\frac{1}{2}(\nabla_i u_k+\nabla_k u_i)
 \quad\quad u_{kk}=\nabla_k u_k.
 \end{eqnarray}
 Hereafter by the repeated indices is denoted summation over all values.
 The trace of the stain tensor, $u_{kk}$,
 is the measure of fraction change in the volume of small elementary mass of solid continuum
 $u_{kk}=\delta {\cal V}/{\cal V}=-\delta {\rho}/{\rho}$ and hence
  \begin{eqnarray}
 \label{e2.3}
 && \delta \rho=- \rho\, u_{kk}\quad u_{kk}=\nabla_k u_k.
 \end{eqnarray}
 The second fundamental statement of elastodynamics is the Newtonian law of viscosity
 which is described by similar tensorial form of constitutive equation expressing
 linear relation  between the viscous stress tensor $\pi_{ik}$ and the rate-of-strain
 tensor ${\dot u}_{ik}$ (e.g. Landau, Lifshits, Kosevich and Pitaevskii 1995)
  \begin{eqnarray}
 \label{e2.4}
  && \pi_{ik}=2 \eta {\dot u}_{ik}+
 \left(\zeta-\frac{2}{3}\eta\right)\,{\dot u}_{ll}\delta_{ik}=2\eta\left({\dot u}_{ik}-\frac{1}{3}
 {\dot u}_{jj}\delta_{ik}\right)+
 \zeta \,{\dot u}_{jj}\delta_{ik}\\
 \label{e2.5}
 &&{\dot u}_{ik}=\frac{1}{2}(\nabla_i {\dot u}_k+\nabla_k {\dot
 u}_i)  \quad\quad {\dot u}_{kk}=\nabla_k {\dot u}_k.
 \end{eqnarray}
 In the analysis of oscillatory behavior of a viscoelastic solid,
 the strength's characteristics of viscoelastic deformations
 -- the shear modulus $\mu$,
 the bulk modulus $\kappa$
 (heaving, both $\mu$ and $\kappa$, physical dimension identical to that for the pressure $P$), the
 shear viscosity $\eta$ and the bulk viscosity $\zeta$ (in general local functions of position)
 are regarded, together with the bulk density $\rho$, as input parameters.
 These are specific to each given sort of viscoelastic
 material whose values are measured in experiments on propagation of material waves
 (so called primary or $P$-waves and secondary or $S$-waves) in an isotropic solid.
 The second law of Newtonian dynamics for an isotropic viscoelastic medium is expressed by
 the Lam\'e-Navier equation
 \begin{eqnarray}
 \label{e2.6}
  \rho{\ddot u}_i=\nabla_k\,p_{ik}+\nabla_k\,\pi_{ik}.
 \end{eqnarray}
 Henceforth the overdot denotes partial differentiation with respect to
 time.
 The energy conservation in the process of viscoelastic distortions
 is controlled by the equation
 \begin{eqnarray}
 \label{e2.8}
 \frac{\partial }{\partial t} \frac{\rho{\dot u}^2}{2}=-
 [p_{ik}+\pi_{ik}]\,{\dot u}_{ik}+ \nabla_k[(p_{ik}+\pi_{ik})\, {\dot u}_i]
 \end{eqnarray}
 which results from (\ref{e2.6}) after multiplication with
 ${\dot u}_i$.

 From above it follows that non-compressional response of a viscoelastic
 solid to short-time applied load, which is not accompanied by local fluctuations in density, $\delta
 \rho=-\rho\,u_{kk}=0$, the tensors of viscoelastic stresses are reduced to
   \begin{eqnarray}
  \label{e2.9}
 p_{ik}=2\mu\,u_{ik}\quad\quad \pi_{ik}=2\eta\,{\dot u}_{ik}\quad\quad u_{kk}=\nabla_k
 u_k=0.
 \end{eqnarray}
 When studying the fast mechanical response of a viscoelastic
 solid by waves of material displacements it is customarily assumed that intrinsic distortions are dominated by
 gradients in the tensor of strains and rate-of-strains
 \begin{eqnarray}
  \label{e2.10}
 \vert u_{ik}\nabla_k{\mu}\vert << \vert \mu\nabla_k\,u_{ik}\vert \quad\quad \vert {\dot u}_{ik} \nabla_k\eta\vert << \vert
 \eta\nabla_k{\dot u}_{ik}\vert
 \end{eqnarray}
 implying
\begin{eqnarray}
  \label{e2.11}
 && f_i^H=\nabla_k p_{ik}=2u_{ik}\nabla_k \mu+2\mu\nabla_k\,u_{ik}\approx \mu\nabla^2 u_i \\
  \label{e2.12}
 && f_i^N=\nabla_k \pi_{ik}=2{\dot u}_{ik}\nabla_k \eta+2\eta\nabla_k\,{\dot u}_{ik}\approx\eta\nabla^2
 {\dot u}_i.
 \end{eqnarray}
 so that parameters of viscoelasticity are regarded as constants.
 In this approximation the basic equation of solid mechanics is reduced to $\rho{\ddot
 u}_i=\mu\,\nabla^2 u_i+\eta\,\nabla^2 {\dot u}_i$.

\subsection{Shear oscillations of a perfectly elastic solid globe in the regime of standing waves of
material displacements}

 To accentuate the subject of our investigation, we pause for a while on
 the case of perfectly elastic ($\eta=0$) solid globe with uniform profile of
 shear modulus $\mu=constant$. In this case equation of elastodynamics is reduced to
 the canonical wave equation $\rho{\ddot {\bf u}}=\mu\,\nabla^2 {\bf u}$
 showing that perfectly elastic solid can transmit compressionless plane-wave perturbation ${\bf
 u}={\bf u}_0 \exp{i(\omega t-{\bf k\,r})}$ by transverse wave ($\nabla\cdot {\bf u}=0\to {\bf k}\cdot {\bf u}=0$)
 of elastic shear traveling with the speed $c_t^2=\mu/\rho$ and characterized by the dispersion free law of
 propagation $\omega^2=c_t^2 k^2$. This suggests the problem of oscillatory response of a perfectly elastic solid
 globe can be solved by working from Helmholtz equation for the standing shear waves $\nabla^2
 {\bf u}+k_t^2{\bf u}=0$. As was mentioned, in this form the problem of
 oscillatory response of a perfectly elastic sphere has been
 posed and solved by Lamb (1882). Specifically, it has been suggested that
 oscillatory modes in an elastic sphere, termed as spheroidal ($s-mode$) and torsional or toroidal ($t-mode$), can be
 unambiguously classified by two general, regular in origin, solutions to the vector
 Helmholtz equation given by  mutually orthogonal and different in parity solenoidal vector fields --
 the poloidal and the toroidal fields of fundamental basis (Chandrasekhar, 1961).
 The fluctuating displacements in spheroidal mode are given by the polar (normal or even parity) poloidal vector field
 ${\bf u}_s=\nabla\times\nabla\times [{\bf r} j_\ell(kr)P_\ell(\cos\theta)]\exp{i(\omega t)}$ and in torsional
  mode by the axial (abnormal or odd parity) toroidal field ${\bf
 u}_t=\nabla\times[{\bf r} j_\ell(kr)P_\ell(\cos\theta)]\exp{i(\omega t)}$.
 Here $j_\ell(kr)$ stands for the spherical Bessel function
 and $P_\ell(\cos\theta)$ for the Legendre polynomial of degree $\ell$.
 Based on this it was shown that eigenfrequencies of spheroidal and toroidal modes
 can be computed from the boundary condition of stress free surface: $n_k\,p_{ik}\vert_{r=R}=0$, with $n_k$ being components of the unit vector normal to the
 globe surface. This condition, applied to each of the above two fields of material displacements, leads to
 the transcendent dispersion equations for spherical Bessel functions and its derivatives,
 $F(j_\ell(z),dj_\ell(z)/dz)\vert_{r=R}=0$ where $z=kr$. The roots $z_{\ell i}=k_{\ell i}R$ of these equations
 uniquely define the frequency $\omega_{\ell i}=c_tk_{\ell
 i}=z_{\ell i}c_t/R$ where the subscript $i$ stands for the node number in overtone of given multipole
 degree $\ell$.

 The oscillatory behavior of a perfectly elastic solid
 sphere in the regime of standing waves has been the objective of extensive study in different areas
 of basic and applied research\footnote{It may be relevant to note that in current investigations of
 technological interest, the model of vibrating solid
 sphere is invoked to explain spectra of ultrafine nanoparticles of ordinary, non-quantum, solids
 (e.g. Hern\'andez-Roses, Picquart, Haro-Poniatowski, Kanehisa, Jouanne and Morhange 2003;
 Duval, Saviot, Mermet and Murray, 2005; Bastrukov and Lai 2005).} and perhaps most thoroughly in geophysical
 context of free oscillations of the Earth (e.g. Lamb 1945; Jeffreys 1970; Lapwood and Usami
 1981; Aki and Richards 2002). In this paper the emphasis is laid on less investigated case
 of the solid elastically deformable globe responding to applied compression free load by local fluctuations
 in shear viscoelastic stresses obeying the conditions $\nabla_k
 p_{ik}=0$ and $\nabla_k \pi_{ik}=0$. This specific case is of particular interest for several
 problems of nuclear collective dynamics and asteroseismology of neutron stars.

\section{Quasistatic oscillations of a viscoelastic solid globe}

 The characteristic feature of the solid globe response by
 small amplitude fluctuations in shear elastic
 $p_{ik}=2\mu\,u_{ik}\neq 0$ and viscous $\pi_{ik}=2\eta\,{\dot u}_{ik}\neq
 0$ stresses obeying the conditions, $\nabla_k p_{ik}=0$ and $\nabla_k
 \pi_{ik}=0$, is that the solenoidal field of material
 displacements is subjected to the vector Laplace equation
 \begin{eqnarray}
 \label{e3.1}
 \nabla^2 {\bf u}({\bf r},t)=0\quad\quad \nabla\cdot{\bf u}({\bf
 r},t)=0.
 \end{eqnarray}
 In the above context, this last equation can be thought of as long-wavelength limit of the vector
 Helmholtz equation for the standing-wave regime: in the limit of extremely long wavelengths
 $\lambda\to\infty$ the wave vector $k=(2\pi/\lambda)\to 0$.

 In order to see what goes on with  the energy in the process of such fluctuations of material stress, consider the integral
 equation of energy balance of viscoelastic shear deformations
 \begin{eqnarray}
 \label{e3.2}
  \frac{\partial }{\partial t}\int \frac{\rho {\dot u}^2}{2}\,d{\cal
  V} &=& -\int p_{ik}{\dot u}_{ik}\,d{\cal V}-\int \pi_{ik}\,{\dot u}_{ik}\,d{\cal
  V}\\
  \label{e3.3}
  &=& -2\int \mu\, u_{ik}{\dot u}_{ik}d{\cal V}-2\int \eta\, {\dot u}_{ik}\,{\dot u}_{ik}\,d{\cal
  V}
   \end{eqnarray}
 which is obtained on integrating (\ref{e2.8}) over the globe volume and taking into
 account the condition of stress free surface implying  $\mu\vert_{r=R}=0$ and
 $\eta\vert_{r=R}=0$. One sees that the integrands in the right hand side of (\ref{e3.3}), defining the work done by
 elastic and viscous stresses are independent of spatial derivatives of neither
 viscoelastic parameters $\mu$ and $\nu$ nor shear strain $u_{ik}$ and rate-of-strain ${\dot
 u}_{ik}$. These latter tensors enter these integrals as
 quadratic combinations of the form $u_{ik}{\dot u}_{ik}$ and ${\dot u}_{ik}{\dot
 u}_{ik}$, that is, the indegrands do not comprise second order spatial derivatives of $u_{i}$
 obeying the Laplace equation (\ref{e3.1}). In the meantime,
 the bulk forces of both the Hookean elastic stress $f_i^H$ and the Newtonian viscous
 stresses $f_i^N$ turn to zero:
 $f_i^H=\nabla_k p_{ik}=\mu\nabla^2 u_i\to 0$ and
 $f_i^N=\nabla_k \pi_{ik}=\eta\nabla^2 {\dot u}_i\to 0$. From this the term quasistatic, that is force-free, shear
 oscillations is derived. This observation suggests that analysis of quasistatic oscillations can be
 performed by working from the equation of energy balance, not form dynamical equation as is the above outlined
 case of an elastic solid globe responding to perturbation by standing waves of elastic shear.
 In doing this we take advantage of the energy variational method which is due to Rayleigh (e.g.
 Lamb 1945; Lapwood and Usami 1981).

\subsection{The energy variational method}

 The point of departure is to use separable representation
 of the vector field of displacements and tensor field of shear deformations
 \begin{eqnarray}
 \label{e3.4}
  u_i({\bf r},t)=a_i({\bf r})\,\alpha(t)\quad
 u_{ik}({\bf r},t)=a_{ik}\,({\bf r})\,\alpha(t)\quad
  a_{ik}=\frac{1}{2}\left(\nabla_i\,a_k+\nabla_k a_i\right)
 \end{eqnarray}
 where $\alpha(t)$ is the temporal amplitude and ${\bf a}({\bf r})$ is the solenoidal vector
 field of instantaneous, time-independent, displacements.
 On substituting (\ref{e3.4}) in (\ref{e3.3}) we arrive at
 the well-familiar equations for the amplitude $\alpha(t)$:
 \begin{eqnarray}
 \label{e3.5}
 &&\frac{\partial {\cal H}}{\partial t}=-2{\cal F}
 \quad\quad{\cal H}=\frac{M{\dot \alpha}^2}{2}+
 \frac{K{\alpha}^2}{2}\quad\quad
 {\cal F}=\frac{D{\dot \alpha}^2}{2}\\
  \label{e3.6}
  && M{\ddot \alpha}+D{\dot \alpha}+K\alpha=0
 \end{eqnarray}
 describing damped harmonic oscillator. The Hamiltonian ${\cal H}$ stands for the total energy of dissipative
 free, normal, vibrations and ${\cal F}$ is the Rayleigh's dissipative function.
 The integral coefficients of inertia ${M}$, stiffness ${K}$ and
 viscous friction ${D}$ are given by
   \begin{eqnarray}
   \label{e3.7}
   M = \int \rho(r)\,a_i\,a_i\,d{\cal V}\quad
      K = 2\,\int\mu(r)\,a_{ik}\,a_{ik}\,d{\cal V}\quad
      D= 2\,\int \eta(r)\,a_{ik}\,a_{ik}\,d{\cal V}.
   \end{eqnarray}
 The solution of (\ref{e3.6}) taken in the form $\alpha(t)=\alpha_0\exp(\lambda t)$
 leads to
  \begin{eqnarray}
 \label{e3.8}
 && \lambda=-\frac{{D}}{2{M}}\pm i\left[\frac{K}{M}-
 \frac{D^2}{4M^2}\right]^{1/2}
 =-\tau^{-1}\pm i\Omega
 \end{eqnarray}
 where $\tau$ is the lifetime and $\Omega$ is the frequency
 of oscillations damped by viscosity. As a result we obtain $\alpha(t)=\alpha_0\exp(-t/\tau)\cos(\Omega
 t)$ with
  \begin{eqnarray}
  \label{e3.9}
  &&\Omega^2=\omega^2\left[1-(\omega\tau)^{-2}\right]\quad
  \omega^2=\frac{K}{M}\quad \tau=\frac{2{M}}{D}.
  \end{eqnarray}
  By $\omega$ is denoted the frequency of the free, non-damped, normal oscillations whose
  timing is determined by the condition $\omega\tau>>1$.
  Thus, to compute the frequency and lifetime of quasistatic oscillations
  one need to specify the fields of instantaneous material displacements ${\bf a}({\bf r})$ entering
  the integral coefficients $M$, $K$ and $D$ of oscillating solid globe.
  The radial profiles of density $\rho(r)$, the shear modulus $\mu(r)$ and shear viscosity
  $\eta(r)$ in the solid globe are regarded as input data of the method.

\subsection{Material displacements in spheroidal and torsional modes of quasistatic oscillations}

 It is clear from above that the fields of instantaneous
 material displacements ${\bf a}({\bf r})$ obey too the vector Laplace equation
 \begin{eqnarray}
 \label{e3.10}
 \nabla^2 {\bf a}({\bf r})=0\quad\quad \nabla\cdot{\bf a}({\bf r})({\bf r})=0.
 \end{eqnarray}
 Adhering to the above Lamb's classification of the vibrational
 eigenmodes of a perfectly elastic solid sphere,
 as spheroidal (shake or $s-mode$) and torsional (twist or $t-mode$),
 in the case under consideration, the eigenmodes of quasistatic
 regime of oscillations can be
 specified by two fundamental solutions to the vector Laplace equation (\ref{e3.10}), built on the general,
 regular in origin, solution of the scalar Laplace equation
 \begin{eqnarray}
 \label{e3.11}
 \nabla^2\,\chi({\bf r})=0\quad\quad \chi({\bf r})= r^\ell\,P_\ell(\cos\theta).
  \end{eqnarray}
 The first one,
 describing instantaneous displacements in spheroidal mode of quasistatic oscillations
 is given by the even parity (polar) poloidal vector field
 \begin{eqnarray}
 \label{e3.12}
 &&{\bf a}_s=\frac{A_s}{\ell+1}
 \nabla\times\nabla\times [{\bf r}\,\chi({\bf r})].
 \end{eqnarray}
 Remarkably,
  \begin{eqnarray}
 \label{e3.13}
 \nabla\times\nabla\times [{\bf
 r}\,r^\ell\,P_\ell(\cos\theta)]=(\ell+1)\nabla\,r^\ell\,P_\ell(\cos\theta)
 \end{eqnarray}
 and, hence, (\ref{e3.12}) can be represented in the following
 equivalent form
 \begin{eqnarray}
 \label{e3.14}
 {\bf a}_s=A_s\,\nabla\,r^\ell\,P_\ell(\cos\theta)
 \end{eqnarray}
 exhibiting irrotational character of quasistatic spheroidal
 oscillations: $\nabla\times {\bf a}_s=0$. For the further purpose we note that the
 velocity field of material displacements
 \begin{eqnarray}
 \label{e3.15}
 \delta {\bf v}_s({\bf r},t)={\dot {\bf u}}_s({\bf r},t)={\bf a}_s({\bf r}){\dot \alpha}(t)
 \end{eqnarray}
 in the dipole overtone of spheroidal mode is uniform and thereby corresponds to
 the center-of-mass motion with no change of intrinsic material stresses both elastic $p_{ik}$ and viscous $\pi_{ik}$.
 This means that poloidal dipole field, ${\bf a}_s(\ell=1)=constant$, describes motions of translation character,
 not oscillatory.

 The second fundamental solution to the vector Laplace equation
 describing instantaneous displacements in the torsional mode is given by
 the odd parity (axial) toroidal vector field
 \begin{eqnarray}
 \label{e3.16}
 {\bf a}_t=A_t\nabla\times [{\bf r}\,\chi({\bf r})]=A_t
 [\nabla\,\chi ({\bf r})\times {\bf r}].
 \end{eqnarray}
 The velocity of material displacements in the toroidal mode $\delta {\bf v}_t({\bf r},t)=
 {\dot {\bf u}}_t({\bf r},t)={\bf a}_t({\bf r}){\dot \alpha}(t)$ can be represented in the well-familiar
 rotational form
 \begin{eqnarray}
 \label{e3.17}
 \delta {\bf v}_t({\bf r},t)=
 [\delta \mbox{\boldmath $\omega$}({\bf r},t)\times {\bf r}]\quad\quad
 \delta \mbox{\boldmath $\omega$}({\bf r},t)=\frac{1}{\ell+1}[\nabla\times \delta {\bf v}_t({\bf
 r},t)]
 \end{eqnarray}
  exhibiting differentially - rotational, vortical, character of torsional
 oscillations  with the angular velocity
 \begin{eqnarray}
 \label{e3.18}
 \delta \mbox{\boldmath $\omega$}({\bf
 r},t)=A_t\,\nabla\,\,r^\ell\,P_\ell(\cos\theta)\,{\dot
 \alpha}(t).
 \end{eqnarray}
 This representation of toroidal solution to the vector Laplace
 equation expresses natural extension of the kinematics
 of rigid-body or uniform rotation to the case of differential
 rotation: the angular velocity is the vector field, not a constant vector.
 The case of rigid-body rotation about polar
 axis $z$
 \begin{eqnarray}
 \label{e3.19}
 \delta {\bf v}=[\delta \mbox{\boldmath $\omega$}\times {\bf r}]\quad\quad
 \delta \mbox{\boldmath $\omega$}=\frac{1}{2}\nabla\times\delta {\bf v}={\rm
 constant}
 \end{eqnarray}
 is the particular dipole ($\ell=1$) case of toroidal field given by (\ref{e3.17}) and (\ref{e3.18}).
 The radial profiles of these fields have no nodes in volume of oscillating globe,
 that is, in the interval $[0< r < R]$. In view of this, the term non-radial shear oscillations
 to this type of oscillatory behavior is applied. The
 intrinsic distortions in a viscoelastic sphere undergoing
 non-radial spheroidal and torsional quasistatic oscillations are shown Fig.1.


 \begin{figure}
\centering{ \includegraphics[width=12cm]{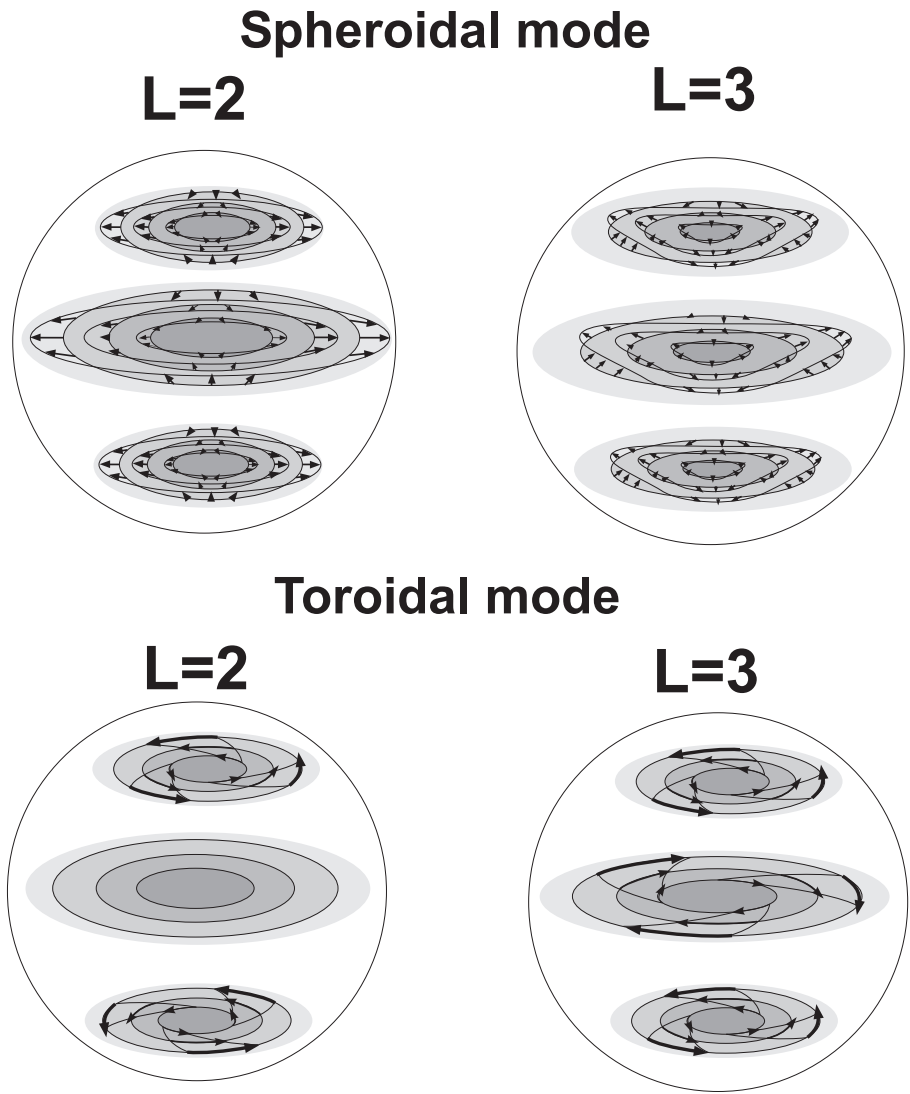}}
 \caption{ Artist view
 of intrinsic distortions in a viscoelastic solid globe
 undergoing quasistatic shear oscillations in quadrupole (L=2) and
 octupole (L=3)
 overtones of spheroidal and toroidal modes.}
\end{figure}


 One of the important features of the energy variational method
 is that it can be used for a wide class
 of problems of different physical interest dealing with non-radial vibrations
 of spherical masses of viscoelastic matter characterized by non-homogeneous distribution of bulk density and
 non-uniform profiles of shear modulus and shear viscosity,
 provided that the fields of induced material displacements in the
 oscillating object obey the vector Laplace equation.

\subsection{Homogeneous solid globe model of uniform density and constant transport
coefficients of viscoelasticity}

 For the solid globe with $\rho={\rm const},\,\mu={\rm
 const},\, \eta={\rm const}$ the parameters of inertia $M$,
 stiffness $K$ and friction $D$ in spheroidal mode of quasistatic oscillations are given by
 \begin{eqnarray}
 \label{e4.1}
 && M_s=4\pi\rho\,A_s^2 R^{2\ell+1}
 \frac{\ell}{2\ell+1}\quad K_s={8\pi}\,\mu\,A_s^2\,R^{2\ell-1}\ell(\ell-1)\\
 \label{e4.2}
 &&  D_s=
 {8\pi}\,\eta\,A_s^2\,R^{2\ell-1}\ell(\ell-1).
 \end{eqnarray}
 One sees, the arbitrary constant
  $A_s$ enter the last equations as $A_s^2$ and, hence, the frequency and lifetime of $s-mode$,
  which are determined by ratios of these parameters, are independent of specific form of
  arbitrary constant $A_s\neq 0$. Equations (\ref{e4.1}) and (\ref{e4.2})
  show that monopole $\ell=0$ vibrations cannot be excited due to
  adopted approximation of non-compressional deformations.
  The dipole field of material displacements describes translational motion of the globe as a whole with
  constant kinetic energy. In this case the integral
  coefficients of both the elastic restoring force, $ K_s$, and
  the damping friction force, $D_s$ vanish.
  Thus, the quadrupole $\ell=2$ overtone of spheroidal of quasistatic oscillations is the lowest one.
  The spectral formulas for the frequency of
  dissipative free oscillations $\omega_s^2=K_s/M_s$ and the time of their viscous damping $\tau_s=2M_s/D_s$
  as functions of multipole degree $\ell$ are given by
  \begin{eqnarray}
  \label{e4.3}
 && \omega_s^2=\omega_0^2\,[2(2\ell+1)(\ell-1)]\quad\quad\quad  \tau_s=\frac{\tau_0}{(2\ell+1)(\ell-1)}\\
  \label{e4.4}
 &&\omega_0^2=\,\frac{\mu}{\rho
 R^2}\quad\quad
 \tau_0=\frac{\rho R^2}{\eta}
 \end{eqnarray}
 where $\omega_0$ is the natural unit of frequency and $\tau_0$ of the lifetime of shear vibrations.

 Computed in a similar fashion the integral parameters of torsional non-radial oscillations
 read
\begin{eqnarray}
 \label{e4.5}
 && M_t=4\pi\rho\, R^{2\ell+3}
 \frac{A_t^2\ell(\ell+1)}{(2\ell+1)(2\ell+3)}\quad
 K_t = {4\pi}\,\mu
 \,R^{2\ell+1}\frac{A_t^2\ell(\ell^2-1)}{2\ell+1} \\
  \label{e4.6}
 &&  D_t= {4\pi}\,\eta\,R^{2\ell+1}\frac{A_t^2\ell(\ell^2-1)}{2\ell+1}.
 \end{eqnarray}
  The eigenfrequency of non-dissipative oscillations $\omega_t^2=K_t/M_t$ and the time of their viscous
  dissipation  $\tau_t=2M_t/D_t$ are given by
  \begin{eqnarray}
  \label{e4.7}
 && \omega_t^2=\omega_0^2 [(2\ell+3)(\ell-1)]\quad\quad\quad \tau_t=\frac{2\tau_0}{(2\ell+3)(\ell-1)}
 \end{eqnarray}
 where $\omega_0$ and $\tau_0$ are the above defined natural units
 of frequency and lifetime of these fundamental modes.
 From spectral formulas (\ref{e4.3}) and (\ref{e4.7}) it follows:
 the larger multipole degree of vibration
 $\ell$ the higher frequency and the less lifetime, as pictured in Fig. 2.


\begin{figure}
\centering{\includegraphics[width=12cm]{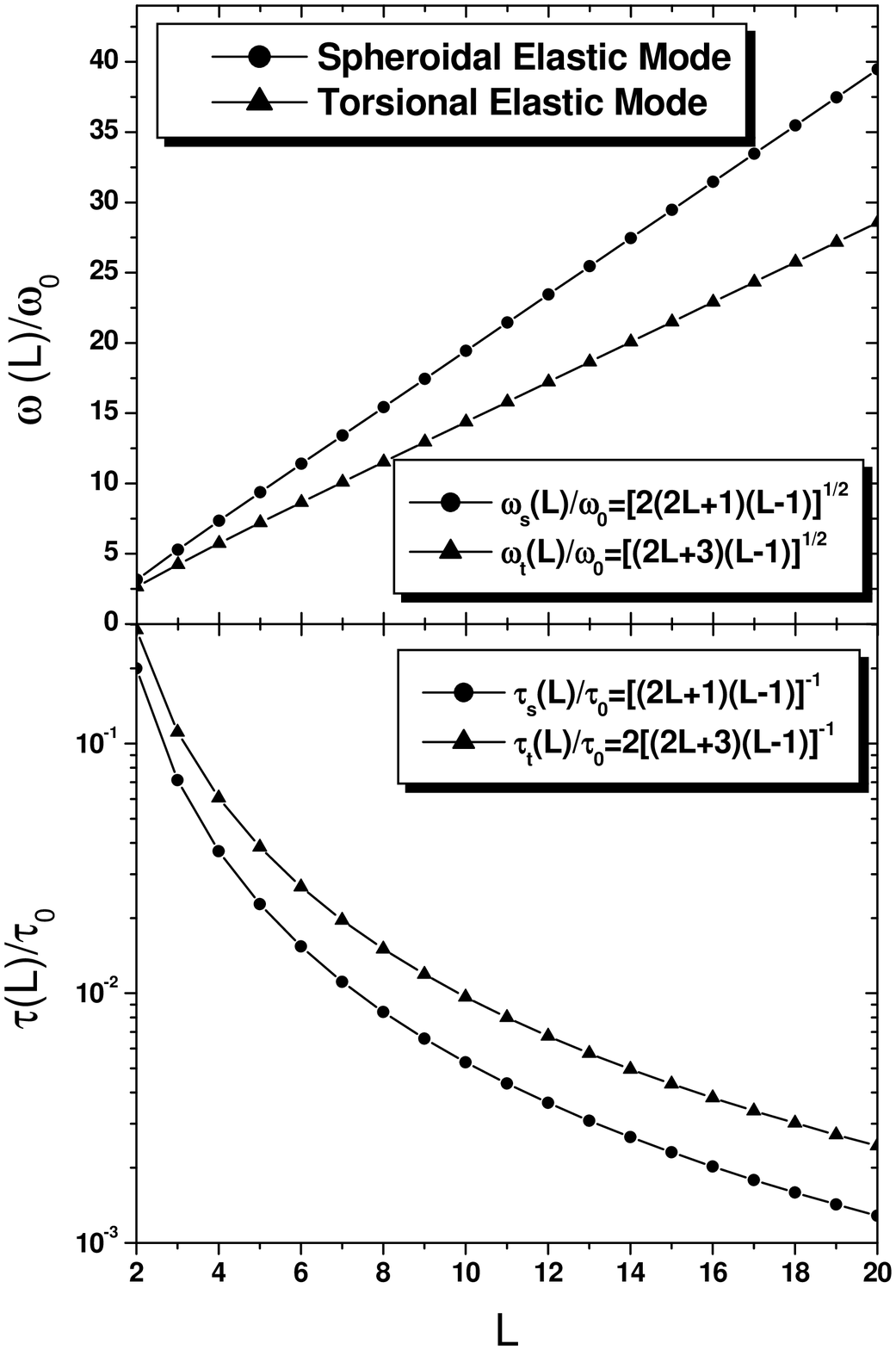}} \caption{
 Frequency and time of viscous damping as functions of multipole
 degree L for quasistatic spheroidal and torsional non-radial
 oscillations of a homogeneous viscoelastic solid globe in natural
 units of the frequency of $\omega_0^2=\mu/(\rho R^2)$ and the lifetime
 $\tau_0=\rho R^2/{\eta}$ of shear vibrations.}
\end{figure}

  Its worth noting that taking $A_t=R^{-(\ell-1)}$ we find that at $\ell=1$  the vector of
  vorticity, $\delta \mbox{\boldmath $\omega$}=\nabla\times \delta {\bf v}={\bf e}_z{\dot
  \alpha}$, is uniform vector directed along polar axis. With this $A_t$, the above computed parameter of
  inertia $M_t$ as a function of multipole
  degree $\ell$ becomes
  \begin{eqnarray}
  \label{e4.8}
  M_t(\ell)=4\pi\rho R^5\frac{\ell(\ell+1)}{(2\ell+1)(2\ell+3)}.
  \end{eqnarray}
  It is easy to see that at $\ell=1$ this parameter equals to the moment of
  inertia of rigid sphere of mass ${\cal M}$:
   $$M_t(\ell=1)=J=\frac{2}{5}{\cal M}R^2\quad\quad {\cal M}=\frac{4\pi}{3}\rho\,R^3. $$
  This is the case when solid globe sets in rigid-body rotation with
  constant kinetic energy. In this case the spring constant $K_t$, and
  parameter of viscous friction, $D_t$, turn to zero and
  the total energy equals the kinetic energy of rigid body rotation: $T=J\omega^2/2$. Thus,
  the torsional quadrupole $\ell=2$ vibration, the twist, is the lowest overtone of the toroidal
  oscillatory mode too.

  It may be noted that frequency of quasistatic oscillations of a perfectly elastic solid globe
  can be derived by staring from quite different footing (Bastrukov
  1994), and as was shown by Zhen Ye (2000) these spectral formulas can be regained by working from solutions
  of the standing wave regime of oscillations by taking in resultant dispersion equations, for both spheroidal
  and toroidal modes, the long wavelength limit. However, this latter approach becomes
  less efficient when shear modulus is a function of position. Also, it is notable
  that the effect of viscosity on the motion of solid material
  continuum is described by the same Newtonian law as in the viscous liquid (Landau, Lifshitz, Kosevich and Pitaevskii 1995).
  Thereby it seems worthy of comparing the above results for the damping time of
  quasistatically oscillating viscoelastic solid globe with that of fluid-mechanical calculation
  for time of viscous damping of non-radial oscillations of a liquid drop.

\subsection{Fluid-mechanical approach to the viscous damping of non-radial oscillations of an incompressible liquid drop}

 The fluid mechanical problem of damping of irrotational oscillation of a spherical mass of an
 incompressible viscous liquid has first been tackled by Lamb (1945) too based on
 the Navier-Stokes equation of hydrodynamics of viscous
 incompressible liquid
\begin{eqnarray}
 \label{e4.9}
 && \rho\,{\delta\dot  v}_i=-\nabla_i\delta p+\nabla_k \delta \pi_{ik}\quad\quad \delta \pi_{ik}=2\eta\,\delta v_{ik}\\
 && \delta v_{ik}=\frac{1}{2}[\nabla_i
 \delta v_k+\nabla_k \delta v_i]\quad \quad \delta v_{kk}=\nabla_k\,\delta v_k=0.
  \end{eqnarray}
 The equivalent vector form of (\ref{e4.9}) reads
\begin{eqnarray}
 \label{e4.9A}
 \rho\delta {\dot {\bf
 v}}=-\nabla \delta p+\eta\nabla^2 \delta {\bf v}\quad\quad \nabla\cdot \delta {\bf
 v}=0\quad \nabla\times \delta {\bf
 v}=0.
 \end{eqnarray}
 Note that irrotational character of the field of velocity $\delta {\bf v}$ in fluctuating
 incompressible flow implies $\delta v_i=-\nabla_i\,\delta \chi$, and
 hence: $\nabla\cdot\delta {\bf v}=0\to\nabla^2 \delta \chi=0$ where $\delta \chi=A_\ell\,r^\ell
 P_\ell(\cos\theta)\,{\dot \alpha}(t)$, so that $\delta {\bf v}$ is the solution to the vector Laplace equation
 $\nabla^2\delta {\bf v}=0$. Understandably that the condition of irrotational flow $(\nabla\times \delta
 {\bf v}=0)$, typical of hydrodynamical consideration, excludes solution of the vector Laplace equation
 describing differential rotational fluctuations. In Lamb's calculations by Rayleigh' method the focus was laid on the non-compressional oscillations
 in which $\delta p=0$. In this case the energy balance equation
\begin{eqnarray}
 \label{e4.10}
 \frac{\partial }{\partial t}\int \frac{\rho\,{\delta {\dot v}}^2}{2}\,d{\cal V}= - \int \delta \pi_{ik}\,\delta v_{ik}
d{\cal V} =-2 \int \eta\, \delta v_{ik}\,\delta v_{ik} d{\cal V}.
  \end{eqnarray}
 with help of substitution $\delta v_i({\bf r},t)=a_i({\bf r})\,{\dot
 \alpha}(t)$ is reduced to
  \begin{eqnarray}
 \label{e4.12}
 &&\frac{d{\cal T}}{dt}=- 2{\cal F}\quad {\cal T}=\frac{M{\dot \alpha}^2}{2}\quad\quad
 {\cal F}=\frac{D{\dot\alpha }^2}{2} \\
 \label{e4.13}
 && M=\int \rho\,a_i\,a_i\,d{\cal V}\quad\quad D= 2\int \eta\,
 a_{ik}\,a_{ik}\,d{\cal V}\quad a_{ik}=\frac{1}{2}(\nabla_i a_k + \nabla_k
 a_i).
  \end{eqnarray}
 where ${\cal T}$ is kinetic energy of oscillations and ${\cal F}$ is
 the dissipative function. Note, the analytic form of inertia $M$ and viscous friction $D$
 is precisely coincides with the above obtained integral parameters of quasistatic oscillations of
 a viscoelastic solid globe. The next step is to represent temporal
 amplitude of oscillations in the form
 \begin{eqnarray}
 \label{e4.14}
 {\alpha}(t)=A(t)\cos \omega t
  \end{eqnarray}
  where $\omega$ is frequency of oscillations and $A(t)$ smoothly varying function of time.
  With this form of $\alpha$ the mean value (averaged over the period of vibration) of
  the kinetic energy and dissipative function are given by $<{\cal T}>=(1/2)\omega^2M\,A^2(t)$ and
  $<{\cal F}>=(1/4)D\,\omega^2\,A^2(t)$, respectively. The key assumption of this line of argument is that
  the equation of energy balance, (\ref{e4.12}), expressing the link between the time-dependent kinetic
  energy ${\cal T}$ and
  doubled dissipative function $2{\cal F}$, preserves its validity for the mean values of these
  quantities
 \begin{eqnarray}
  \label{e4.15}
 && \frac{d}{dt}<{\cal T}>=-2<{\cal F}>\quad\to\quad  M\,\frac{d{A}(t)}{dt}=
 -\frac{1}{2}D\,A(t)\\
  \label{e4.16}
 && A(t)=A_0\exp(-t/\tau)\quad \tau=\frac{2M}{D}.
  \end{eqnarray}
 Thus, the computational formula for the
 viscous damping time $\tau$ is identical to that obtained for a viscoelastic solid
 globe. This explains why the spectral formula
 for the time of viscous damping of irrotational quasistatic spheroidal oscillations of the
 solid globe exactly coincides with the Lamb's formula
 for the damping time of the liquid drop oscillations.
 In the context of viscous damping of global non-radial pulsations of neutron
 stars, the above spectral formulas for lifetime of spheroidal
 and toroidal modes are discussed in
 (Cutler and Lindblom 1987; Bastrukov, Weber and Podgainy 1999; Bastrukov, Yang, Podgainy and  Weber 2003).

\subsection{Nuclear giant resonances in the model of quasistatic
oscillations of viscoelastic solid globe}

 The physical idea underlying the nuclear solid globe model is that the order encoded in the shell-model treatment
 of nuclear structure signifies quantum solidification of degenerate Fermi-system of nucleons.
 In this model a nucleus is thought of as an ultrafine
 (femtometer dimension) particle of Fermi-solid -- the quantum (fermionic) condensed
 matter endowed with material properties of viscoelasticity
 typical of ordinary solids. As was mentioned, the credit for recognition of solid mechanical elasticity
 in the nuclear giant-resonant response goes to seminal work of Bertsch (1974).
 The basic inference of this influential paper is that experimentally established smooth
 dependence of the energy of isoscalar giant quadrupole resonance can be understood as manifestation
 of spheroidal mode of non-compressional, that is pure shear, quasistatic oscillations of an ultrasmall spherical
 particle of an elastic Fermi-solid whose shear modulus equals to the pressure of Fermi-degeneracy:
 $\mu=P_F=\rho v_F^2/5$ (see also, Semenko 1977; Nix and Sierk 1980; Wong and Azziz 1981; S. Mi\c sicu 2006).
 By now there are quite cogent arguments
 showing that both the giant electric and magnetic
 resonances can be treated on equal solid-mechanical footing as manifestation of spheroidal (electric) and
 torsional (magnetic) oscillations of nuclear femtoparticle (e.g. Bastrukov, Mi\c sicu and  Sushkov 1993).
 The electromagnetic nomenclature of nuclear giant resonances owe its origin to the type of
 induced electromagnetic moment of fluctuating current
 density: the giant-resonance excitations of electric type are associated with
 spheroidal mode of quasistatic oscillations of irrotational field
 of material displacements ${\bf u}_s$ and those for the magnetic type with torsional mode of quasistatic
 differentially-rotational
 oscillations of toroidal field of material displacements ${\bf u}_t$.
 The multipole degree $\lambda$ of electric ${\cal M}(E\lambda)$ and magnetic ${\cal M}(M\lambda)$ moments
 equal to multipole degree $\ell$ of excited spheroidal and torsional oscillatory
 state, respectively\footnote{The
 electric and magnetic multipole moment of electric current density are given by
\begin{eqnarray}
  \nonumber
 {\cal M}(E\ell)=N_{E\ell} \int {\bf j}({\bf r},t)\cdot \nabla
 r^\ell\,
 P_{\ell}\,d{\cal V}\quad
 {\cal M}(M\ell)=N_{M\ell} \int {\bf
 j}({\bf r},t)\cdot [{\bf r}\times\nabla ]\,r^\ell\, P_{\ell}\,d{\cal V}
 \end{eqnarray}
  where ${\bf j}({\bf r},t)=\rho_e\,{\dot {\bf u}}({\bf r},t)$ with $\rho_e=(Z/A)n$ is the charge density and
  $n$ being the nucleon density. The exact expressions of $N_{E\ell}$ and $N_{M\ell}$
  can be found elsewhere (e.g. Alder and Steffen 1975; Greiner and Maruhn 1996). Note, the quasistatic oscillations of irrotational field of
  material displacements in spheroidal mode ${\bf u}={\bf u}_s={\bf a}_s({\bf r})\,\alpha(t)$ result in excitations of
  vibrational states with non-zero electric multipole moment, whereas magnetic multipole moment for this kind of oscillations is zero.
  The opposite situation takes place for quasistatic oscillations of differentially-rotational displacements
  ${\bf u}={\bf u}_t={\bf a}_t({\bf r})\,\alpha(t)$ in torsional mode which lead to vibrational excited states with
  non-zero magnetic multipole moment, while the electric multipole moment is zero.
  The computational formulas for the total excitation strength (probability) of giant electric
  and magnetic resonance are given by  $B(E\ell)=(2\ell+1)<|{\cal M}(E\ell)|^2>$ and
  $B(M\ell)=(2\ell+1)<|{\cal M}(M\ell)|^2>$, respectively,
  where bracket stands for time average.}.

\begin{figure}
\centering{\includegraphics[width=16cm]{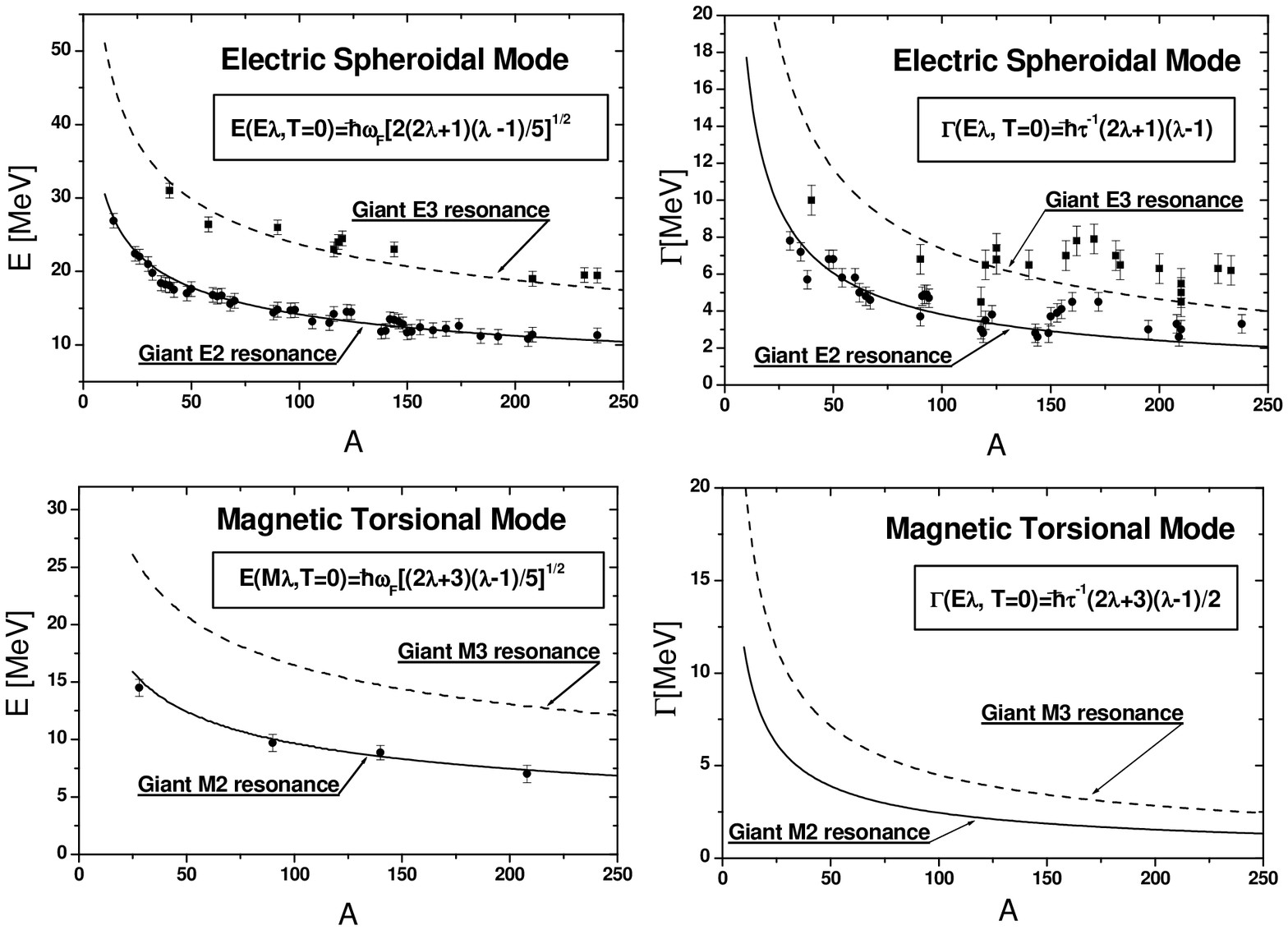}}
  \caption{The energy $E=\hbar\omega$ and width
   $\Gamma=\hbar \tau^{-1}$ of isoscalar electric and magnetic resonances,
   interpreted from equal footing as manifestation of
   spheroidal (electric) and torsional (magnetic) quasistatic
   oscillations whose frequency and lifetime is given by
   (\ref{e4.3}) and (\ref{e4.7}), respectively. The
   multipole degree $\lambda$ of excited electromagnetic moment of the electric current density equal to multipole
   degree of oscillations $\lambda=\ell\geq 2$, and $\omega_F=v_F/R$,
   as pointed out in the text.}
  \end{figure}

 Fig. 3 shows how the spectral formulas (\ref{e4.3}) and (\ref{e4.7}) of the
 nuclear solid globe model can be utilized to extract numerical estimates for the shear
 modulus and shear viscosity of nuclear matter from the experimental data of nuclear physics.
 The nuclear density $\rho=2.8\,10^{14}$ g cm$^{-3}$ and the nucleus radius $R=r_0\,A^{1/3}$ ($r_0=1.2\, 10^{-13}$ cm)
 are well defined quantities. Therefore by fitting the data on
 general trends in the energy centroids and spread widths of isoscalar (with isospin quantum number $T=0$)
 giant resonances, which are computed with use of
 standard quantum mechanical equations for the energy centroid $E=\hbar\omega$ and spread width $\Gamma=\hbar/{\tau}$,
 we can extract the values of two unknown quantities -- the shear modulus $\mu$ and shear viscosity $\eta$ of the
 isospin-symmetric nuclear matter, as was mentioned with $T=0$.
 Note, taking into account that in this model $\mu=(1/5)\rho v_F^2$,
 the natural unit of frequency of shear elastic oscillations $\omega_0$ carrying information on shear modulus
 can be expressed in terms of the Fermi-frequency $\omega_F=v_F/R$
 as follows $\omega_0^2=\mu/(\rho R^2)=(1/5)\omega_F^2$. It is to be emphasized that in the presented in Fig. 3
 comparison
 of theoretical calculations (lines) with data (symbols), the transport coefficients of solid-mechanical
 elasticity $\mu$ and $\eta$
 have been fixed so as to reproduce overall trends in data on the energy centroid and the spread width
 of solely one type of resonance -- the giant electric quadrupole $(E2)$ resonance associated with
 quadrupole overtone of spheroidal quasistatic oscillations. The predictive power
 of the solid globe model is demonstrated by fairly accurate account of
 experimental systematics on the electric octupole ($E3$)
 and the magnetic quadrupole $(M2)$ giant resonances which is attained with no use of any adjustable constants.
\begin{figure}
\centering{\includegraphics[width=12cm]{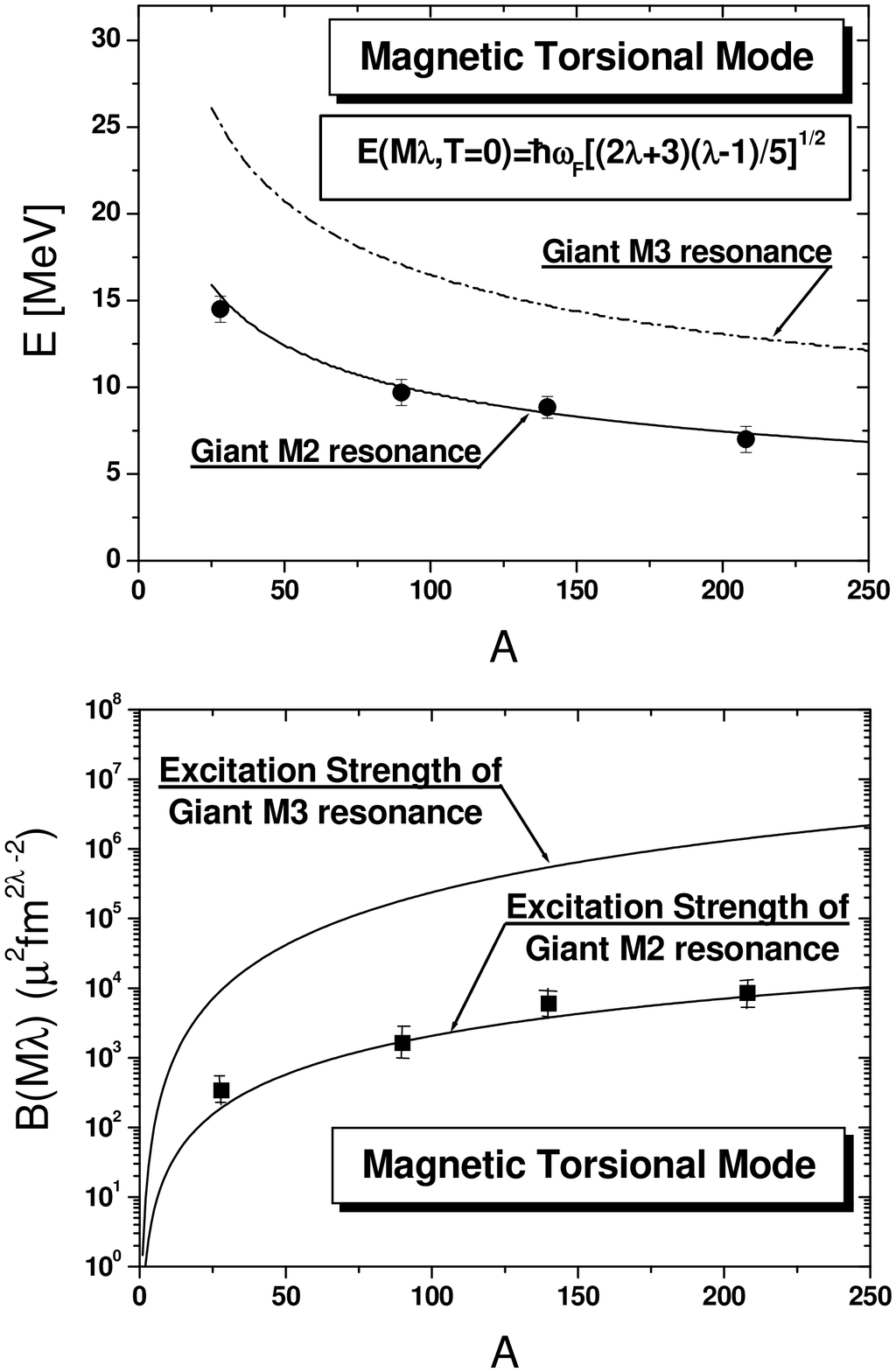}}
  \caption{
   Theoretical calculation (lines) for general trends throughout
   periodic table of the energy centroid and the total excitation
   probability (computed with no adjustable parameters) for the
   nuclear giant magnetic resonances of multipole degree $\lambda\geq 2$ treated in the from stand point
   of the Fermi-solid globe model
   as manifestation of torsional quasistatic oscillations versus
   experimental data (symbols) of S-DALINAC data (Luttge, von
   Neumann-Cosel, Neumeyer and Richter 1996).}
  \end{figure}

 As was mentioned, the giant magnetic resonances exhibiting torsional oscillations of nucleus
 is the mode of giant-resonance nuclear response
 that can be understood only on the basis of the solid globe model, not a liquid drop. One more demonstration of the predictive power of the nuclear Fermi-solid globe model is pictured in Fig.4
 showing adequate description of the electromagnetic excitation strength (total excitation probability) -- $B(M2)$
 of giant magnetic resonance versus mass number $A$ which is attained with no use of any adjustable constants
 (Bastrukov, Molodtsova and Shilov 1995). Also, its worth emphasizing that the spectral formulas
 for the quasistatic shear oscillations match much better the experimental data as compared to
 calculation implying oscillations in the regime of standing waves, extensively studied in (Wong and Azziz 1981).

 The above procedure of extracting the values of transport coefficient of solid-mechanical
 viscoelasticity of nuclear matter -- the
 shear modulus $\mu$ and the shear viscosity $\eta$ yields:
 $$\mu=4\cdot 10^{33}\,\mbox{dyn cm}^{-2} \quad\quad \eta=3\cdot 10^{11}\,\mbox{dyn
 sec} cm^{-2}$$
 (see also Hasse 1978; Nix and Sierk 1980; Kolomietz 1990).
 For comparison, the density of steel
 $\rho\sim 10$ g cm$^{-3}$
 and the shear modulus $\mu\sim 10^{11}$ dyn cm$^{-2}$ (e.g. Fetter and Walecka 1980). In theoretical geophysics
 this last estimate is
 of the same order as the average value for the shear modulus of the Earth's material (e.g. Lapwood and Usami 1981).
 It is appropriate mention here that obtained in this manner estimate
 for the shear viscosity is much less than that
 derived in (Flowers, Itoh 1976: $\eta\sim 10^{18}$ dyn sec cm$^{-2}$) by use
 of absolutely different theoretical arguments
 (see also Cutler and Lindblom 1987; McDermott, Van Horn and Hansen 1988; Bastrukov, Yang, Podgainy  and Weber 2003).
 This inference of nuclear physics is crucial to the asteroseismology of neutron stars studying the timing of oscillatory
 behavior of pulsars and magnetars. With the obtained from giant resonances value of shear viscosity $\eta$,
 one finds that the time of viscous damping of shear oscillations, $\tau_0=\rho R^2/\eta$, in
 the object of nuclear density $\rho$ and radius $R\sim 10^6$ cm, typical of neutron stars, is evaluated
 as $\tau_0\sim 10^7 - 10^8$ years. The net outcome of this finding is that neutron
 stars are expected to be in the restless state of background shear vibrations (triggered either by supernova event
 or starquake) during all its life as a pulsar. As is shown in the following section this inference
 is in line with astrophysical observations and conclusions of several independent theoretical
 investigations on the oscillatory behavior of neutron stars.

\section{The solid globe models of astrophysical interest}

 The above homogeneous models of viscoelastic solid
 globe and viscous liquid drop are of practical usefulness when there are physical arguments justifying
 approximation of uniform profiles for the density and the transport coefficients of shear
 viscoelasticity. This approximation surely becomes unwarranted when studying
 global vibrations of a highly stratified in density solid mass, as is the case
 of solid Earth-like planets and collapsed degenerate stars, like white dwarfs and
 pulsars.
 In this case the shear modulus and shear viscosity cannot be
 considered as having constant values throughout the star volume
 (e.g. Jeffreys 1976; Lapwood and Usami 1981). It is the purpose of
 this section to
 investigate quasistatic oscillations in several mathematical models of viscoelastic solid globe
 with non-uniform profiles of density and parameters viscoelasticity.

 Bearing in mind the astrophysical context of the models under consideration, in what follows we discuss
 quasistatic global shear vibrations of self-gravitating masses whose
 state of gravitational equilibrium is determined by coupled equations for
 the potential of self-gravity and equation for the pressure
 \begin{eqnarray}
 \label{e5.1}
 \nabla^2 \Phi(r)=-4\pi G \rho(r) \quad \quad \nabla
 P(r)=\rho(r)\,\nabla
 \Phi(r).
  \end{eqnarray}
 which, of course, valid for both liquid and solid stars, by virtue of universal character
 of Newtonian law of gravitation. In these equations, the density profile $\rho(r)$ is regarded
 as a given function of distance from the star center. Then having solved the first equation in this line
 for the potential of self-gravity $\Phi(r)$ we can obtain the
 profile of pressure $P(r)$ providing gravitational stability of the
 star. In doing this the standard boundary
 conditions of gravitostatics for the potential on the globe surface, $\Phi_{in}=\Phi_{out}\vert_{r=R}$
 and $\nabla_r \Phi_{in}=\nabla_r \Phi_{out}\vert_{r=R}$, and the condition of stress-free surface
  for the pressure, $P_{r=R}=0$, should be used.

  As a representative example, for the standard model of homogeneous star with $\rho=constant$ throughout
  the star volume, from
  (\ref{e5.1}) it follows that profile of pressure inside the star
  is given by $P(r)=P_c[1-(r/R)^2]$, where $P_c=(2\pi/3)G\rho^2\,R^2$ is the pressure in the star
  center. On the other hand, it should be
  clearly realized, the central pressure must be evaluated
  by use of physical arguments regarding the equation of state of stellar matter and dynamical processes
  opposing the self gravity.
  For instance, in a highly idealized model of neutron
  star, thought of as a homogeneous spherical mass of self-gravitating degenerate neutron matter
  condensed to the normal nuclear density, the central pressure counterbalancing
  the self-gravity is taken in the form of
  equation of state for  degenerate Fermi-gas of non-relativistic neutrons $P_F=K\rho^{5/3}$.
  Then, equating the central pressure $P_c=(2\pi/3)G\rho^2\,R^2$ obtained from the equation of gravitational
  equilibrium (\ref{e5.1})
  with $P_F$ we obtain the equation for the neutron star radius $R$ which leads to canonical
  estimates $R\approx 12$ km and mass $M=(4\pi/3)\rho R^3\approx 1.3\,
  M_\odot$. This is perhaps the most simple example showing that
  the neutron star radius is primarily determined by the behavior of degeneracy pressure of the neutron star matter
  in the vicinity of the nuclear saturation density (e.g. Lattimer and Prakash 2001).
  Also, it may be worthy of noting that from quantum-mechanical side, the fact that
  the pressure in the star counterbalancing Newtonian self-gravity is determined by
  the degeneracy pressure of Fermi-gas of independent neutron-like quasiparticles means that
  the single-particle states of incessant quantum-wave Fermi motion of an individual quasiparticle in the mean field
  of self-gravity are described by Schr\"odinger equation coupled with Poisson
  equation for the potential of the mean gravitational field
  $\Phi(r)$ in the star
  \begin{eqnarray}
  \nonumber
\label{b1.1} &&
 i\hbar\frac{\partial \psi}{\partial t}=\left[-
  \frac{\hbar^2}{2m_n}\nabla^2+U(r)\right]\psi\quad\quad
  U(r)=-m_n\Phi(r)\\
 \label{b1.2}
 &&  \nabla^2 \Phi(r)=-4\pi G \rho(r) \quad\quad
 \rho=\frac{m_n}{3\pi^2}\,k_F^3=2.8\times 10^{14}\,{\rm g \,cm^{-3}}
  \end{eqnarray}
 where $U$ is the potential energy of the in-medium neutron quasiparticle in its
 unceasing Fermi-motion whose collision free character is provided by Pauli exclusion principle;
  $\rho$ stands for the uniform density in the star bulk equal to the normal nuclear density and
 $m_n$ is the effective mass of the neutron quasiparticle.
 The  solution to (\ref{b1.2}) having the form
 $\Phi(r)=(2\pi/3)\,G\rho\,(3R^2-r^2)$
 suggests that the potential energy $U$ can be represented in the well-familiar form of
 spherical harmonic oscillator
  \begin{eqnarray}
 \label{b.4}
 U=-m_n\,\Phi(r)=-U_G^0+\frac{m\,\omega_G^2\,r^2}{2}\quad\quad U_G^0=2\pi\,m\,\rho\,G\,R^2\quad\quad
 \omega_G^2=\frac{4\pi}{3}\,G\rho
  \end{eqnarray}
where  $U_G^0$ is the depth of spherical gravitational trap and
$\omega_G$ is the characteristic unit of frequency of
gravitational vibrations. The stationary states of single-particle
Fermi-motion of individual neutron in
 the mean field of self-gravity are described by Hamiltonian
 \begin{eqnarray}
  \label{b.6}
  H\psi=\left[-\frac{\hbar^2}{2m}\nabla^2+\frac{m\,\omega_{G}^2\,r^2}{2}
 \right]\psi=\epsilon\psi\quad\epsilon=E+U_{G}
 \end{eqnarray}
 whose spectrum of energies, accounted from the bottom of potential well, is well-known
 $\epsilon_N=\epsilon_{0}(N+3/2)$ where $\epsilon_0=\hbar\omega_G\approx 10^{-11}$\,eV is
 the energy of zero-point oscillations which is the measure of the energy distance between single-particle
 states of neutron quasiparticles in the potential of self-gravity of neutron star (note the average
 distance between discrete states
 of degenerate electrons in terrestrial solids such as metals and semiconductors $\Delta\sim 10^{-18}-10^{-20}$\,eV).
 Here  $N=2\ell+n$ is the shell's quantum number of harmonic oscillator, $n$
 and ${\ell}$ are the principle and orbital quantum numbers of
 single-particle orbits, respectively. This shows that the shell-ordered clusterization of
 single-particle energies of neutron quasiparticles in the potential of mean gravitational
 field of the neutron star model has common features with that for nucleons in the mean field
 potential of the nuclear shell model having different physical origin.
 Thus, in addition to the similarity in nucleon-dominated composition and the fact of common genetic
 origin of nuclei and neutron stars rooted in supernovae of second type, an identical
 quantum-mechanical mean field treatment of orderly-organized intrinsic single-particle structure of these objects
 indicates that they are made 
 of the identical in its  material properties degenerate Fermi-matter endowed with solid-mechanical
 property of viscoelasticity -- elastically deformable Fermi-solid.

  One more cognitive value of this historically first and admittedly
  idealized homogeneous model (using the only one input parameter -- the density of nuclear
  matter) is in highlighting the compressional effect of gravity, that is, in showing
  that material pressure is increased in the direction to the star center whereas the density of stellar matter
  remains constant (the well-familiar example is pressure increase
  with the depth of ocean whereas the density of water is practically unaffected).

  Bearing this in mind, it seems not inconsistent to assume that compressional effect of gravity on the
  viscoelastic parameters of self-gravitating solid mass is analogous to that for the pressure
  $P$.  More specifically, it is assumed that the shear modulus profile
  $\mu=\mu(r)$ (having physical dimension of pressure $P$) and the shear modulus profile $\eta=\eta(r)$ are
  identical to that for the pressure profile $P(r)$. Based on this conjecture, in the reminder of this paper
  we compute the frequency and lifetime of quasistatic oscillations for
  several models  of astrophysical interest by evaluating first the radial profile of
  pressure from equations of gravitational equilibrium (\ref{e5.1}).

\subsection{Solid star model with uniform density and non-uniform profiles of shear modulus and shear viscosity}

 Following the above line of argument, consider first the model of uniform density
 and non-uniform shear modulus and shear viscosity whose profiles are taken to be identical to that
 for the pressure of self-gravity:
   \begin{eqnarray}
   \label{e5.2}
 \rho={\rm constant}\quad \mu(r)=\mu_c
    \left[1-\left(\frac{r}{R}\right)^2\right]\quad
  \eta(r)=\eta_c\left[1-\left(\frac{r}{R}\right)^2\right]
   \end{eqnarray}
   where $\mu_c$ and $\nu_c$ the shear modulus
   and shear viscosity of stellar matter in the star center, respectively.
   Then, for integral parameters of global quasistatic oscillations in spheroidal mode we obtain
 \begin{eqnarray}
 \label{e5.3}
 && M_s=4\pi\rho\,A_s^2
 R^{2\ell+1}
 \frac{\ell}{2\ell+1}\quad K_s =
 16 \pi\mu_c  A_s^2\,R^{2\ell-1}\frac{\ell(\ell-1)}{(2\ell+1)}\\
 \label{e5.4}
 && D_s=
 16 \pi\eta_c  A_s^2
 R^{2\ell-1}\frac{\ell(\ell-1)}{(2\ell+1)}
  \end{eqnarray}
 and for the frequency and lifetime we get
 \begin{eqnarray}
  \label{e5.5}
 && \omega_s^2=4\omega_0^2(\ell-1)\quad
    \tau_s=\frac{8\tau_0}{(\ell-1)}\quad \omega_0^2=\,\frac{\mu_c}{\rho
 R^2}\quad
 \tau_0=\frac{\rho R^2}{\eta_c}.
  \end{eqnarray}
The integral parameters of quasistatic differentially rotational
oscillations are given by
 \begin{eqnarray}
 \label{e5.6}
 && M_t=4\pi\rho\,A_t^2 R^{2\ell+3}
 \frac{\ell(\ell+1)}{(2\ell+1)(2\ell+3)}\\
 \label{e5.7}
 && K_t=8\pi\mu_c R^{2\ell+1}A_t^2
 \frac{\ell(\ell^2-1)}{(2\ell+1)(2\ell+3)}\\
 \label{e5.8}
 && D_t =8\,\pi\,\eta_c\,R^{2\ell+1}\,A_t^2\,\frac{\ell(\ell^2-1)}{(2\ell+1)(2\ell+3)}
 \end{eqnarray}
 and for the frequency and lifetime of toroidal mode we obtain
 \begin{eqnarray}
 \label{e5.9}
 \omega_t^2=2\omega_0^2(\ell-1)\quad \quad
 \tau_t=\frac{4\tau_0}{(\ell-1)}.
 \end{eqnarray}
 Thus, the compressional effect of gravity on transport
 coefficients of solid-mechanical viscoelasticity of stellar matter
 is manifested in the spectral formulas for the frequency and
 lifetime of global oscillations. Specifically, for $\mu_c$ and
 $\eta_c$ taken equal to $\mu$ and $\eta$ of the homogeneous model,
 we find that the frequency of quasistatic global oscillations of
 the non-homogeneous solid globe is lower than that for homogeneous
 one and, hence, period of oscillations, $P=2\pi/\omega$, is
 longer. The lifetime of such oscillations in a non-homogeneous solid
 globe are longer as compared to that in the model with homogeneous
 density and uniform profiles of parameters of viscoelasticity.

 Fig.5 shows that periods of background elastic pulsations of
 a homogeneous neutron star model
 computed with use of spectral formulas (\ref{e5.5}) and
 (\ref{e5.9}) with the value of shear modulus extracted from data on nuclear giant resonances
 coincides with the timing of microspikes in the windows between the main
 pulses. Taking into account the above inference regarding the damping time
 of global quasistatic elastic vibrations of pulsars by viscosity of neutron star matter
 one can conclude that these microspikes owe its origin to non-radial elastodynamic (EDM)
 spheroidal and torsional pulsations triggered by neutron star quakes (e.g. Van Horn 1980, McDermott,
 Van Horn and Hansen 1988; Bastrukov Weber and Podgainy 1999).

\begin{figure}
\centering{\includegraphics[width=12cm]{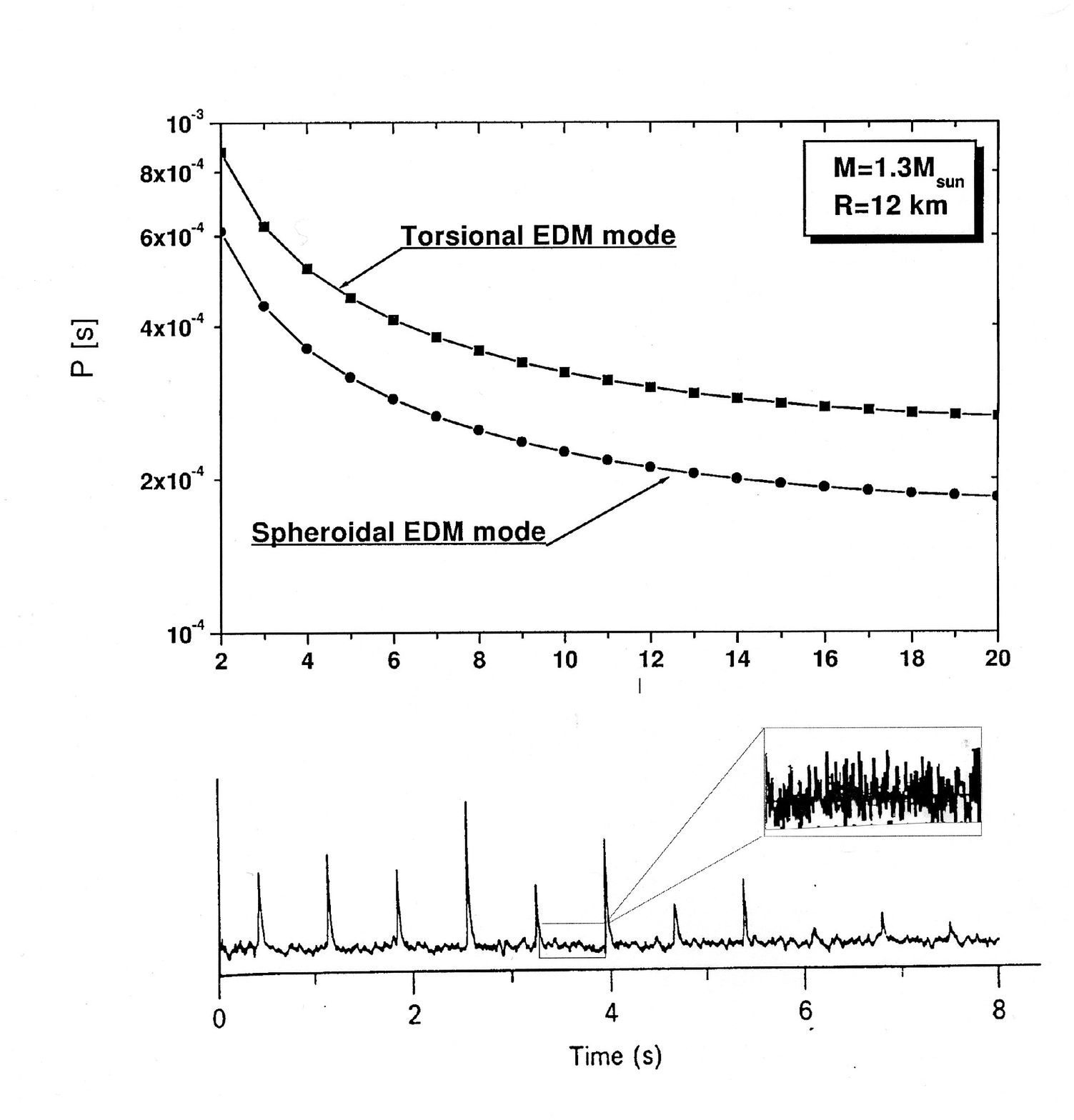}}
\caption{Periods P against multipole degree L for
 spheroidal and torsional elastodynamic (EDM) modes of background
 non-radial pulsations of neutron star manifested by
 microspikes of millisecond duration in the windows of the rotation driven
 main pulse train.}
\end{figure}

 Deserved for special comment is the following particular case. As was
 mentioned, the physical dimension of shear modulus $\mu$ is the
 same as of pressure $P$. Bearing this in mind, consider the model of homogeneous
 solid star putting
 \begin{eqnarray}
 \label{e5.10}
  \rho={constant}\quad\quad \mu(r)=\mu_c
    \left[1-\left(\frac{r}{R}\right)^2\right]\quad
    \mu_c=P_c=(2\pi/3)G\rho^2\,R^2.
 \end{eqnarray}
 In terms of this last parametrization of $\mu_c$, the eigenfrequencies of
 global non-radial spheroidal and torsional quasistatic
 oscillations are represented as follows (Bastrukov 1993, 1996)
 \begin{eqnarray}
 \label{e5.11}
  \omega^2_s=2\omega_G^2(\ell-1)\quad\quad \omega_t^2=\omega_G^2(\ell-1)\quad
 \quad \omega_G^2=\frac{4\pi}{3}\rho\,G.
 \end{eqnarray}
 For a solid star -- self-gravitating mass of a viscoelastic
 solid, the last spectral formulas
 seems to have the same physical meaning as Kelvin's formula does
 \begin{eqnarray}
 \label{e5.12}
 && \omega^2_f=\omega_G^2\frac{2\ell(\ell-1)}{2\ell+1}\quad
 \quad \omega_G^2=\frac{4\pi}{3}\rho\,G
 \end{eqnarray}
 for a liquid star --  self-gravitating mass of an incompressible inviscid
 liquid. This last spectral formula can be obtained by the above expanded
 method  from equations of canonical hydrodynamical model for a self-gravitating
 inviscid liquid relying on equations
 of fluid mechanics coupled with the equation of self-gravity
 \begin{eqnarray}
 \label{K.1}
 \frac{d\rho}{dt}=-\rho
 \frac{\partial v_k}{\partial x_k}\quad\quad
 \rho\frac{dv_i}{dt}=-
 \frac{\partial p}{\partial x_i}+\rho\frac{\partial \Phi}{\partial
 x_i}\quad\quad \nabla^2 \Phi=-4\pi G\rho.
 \end{eqnarray}
 (e.g. Lamb 1945; Chandrasekhar 1961; Bastrukov 1996).
 The comparison of (\ref{e5.11}) and (\ref{e5.12}) is depicted in Fig.6.

 \begin{figure}
\centering{\includegraphics[width=12cm]{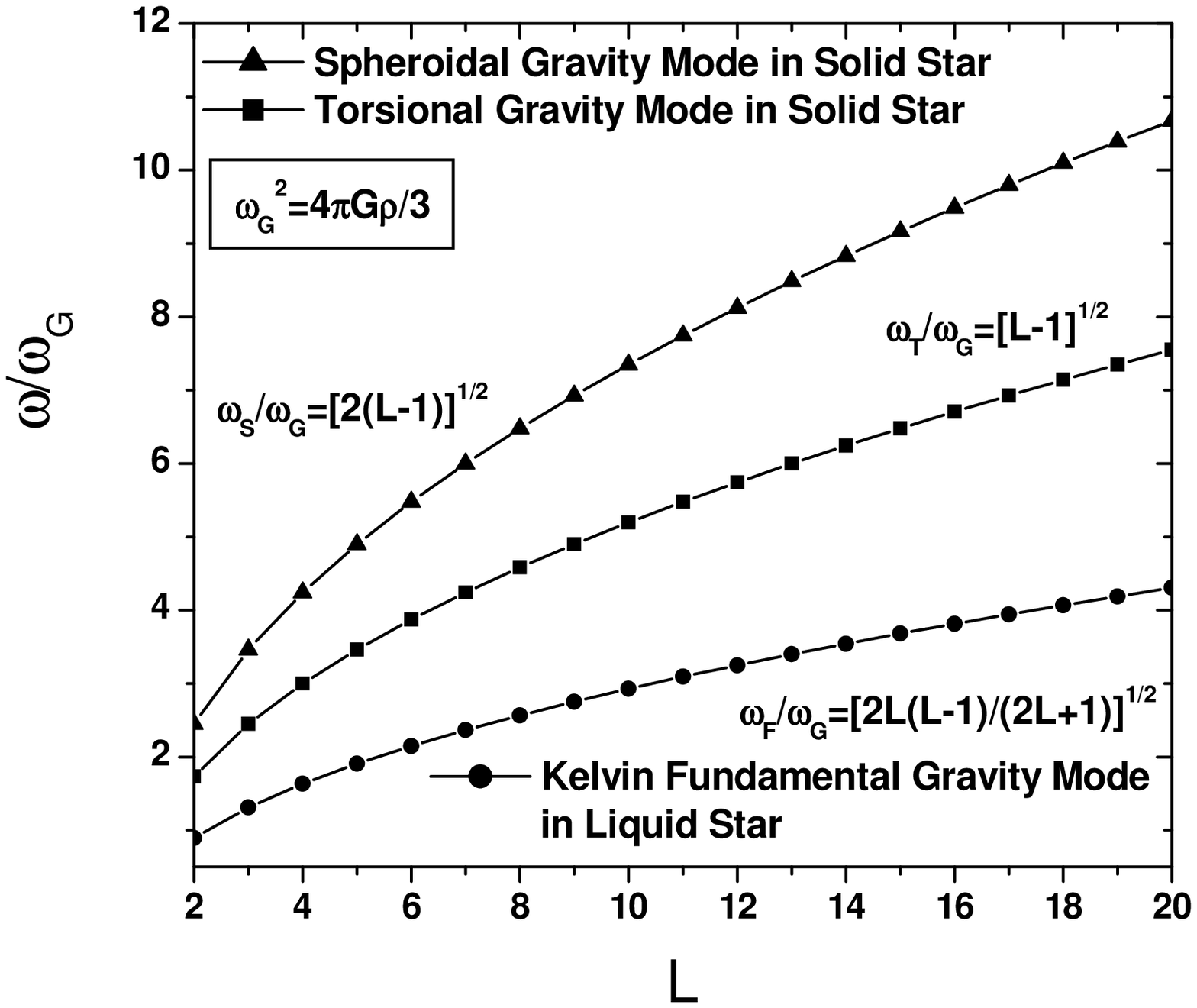}}
 \caption{Fractional frequency $\omega/\omega_G$ versus multipole
 degree L of non-radial oscillations for fundamental spheroidal and
 toroidal gravity modes in a solid star in juxtaposition with that
 for the Kelvin's fundamental gravity mode in a liquid star.}
 \end{figure}

\subsection{Solid star model with the non-uniform density and non-uniform profiles of shear modulus and shear viscosity}

 Consider a solid star model with the non-uniform distribution of density
 \begin{eqnarray}
 \label{e5.20}
 && \rho(r)=\rho_c\left[1-\left(\frac{r}{R}\right)^2\right].
 \end{eqnarray}
 After long but simple calculation of pressure from equations of gravitational
 equilibrium (\ref{e5.1}) we obtain
 \begin{eqnarray}
 \label{e5.21}
 &&
 P(r)=P_c\left[1-\left(\frac{r}{R}\right)^2\right]^2\left(1-\frac{r^2}{2R^2}\right)\quad\quad
 P_c=\frac{4}{15}\pi\,G\,\rho_c^2\,R^2.
 \end{eqnarray}
 Taking into account the above conjecture of similarity in form of radial profiles for
 viscoelastic parameters and pressure we put
\begin{eqnarray}
 \label{e5.22}
 \mu(r)=\mu_c\left[1-\left(\frac{r}{R}\right)^2\right]^2\left(1-\frac{r^2}{2R^2}\right)\,\,
 \eta(r)=\eta_c\left[1-\left(\frac{r}{R}\right)^2\right]^2\left(1-\frac{r^2}{2R^2}\right)
 \end{eqnarray}
 where $\mu_c$ and $\nu_c$ are the sear modulus and shear viscosity in the globe center.
 Omitting tedious calculation of integrals for $M$, $K$ and $D$, which has been performed with
 help of results presented in appendices, we obtain
 \begin{eqnarray}
 \label{e5.23}
 && \omega_s^2=\omega_0^2\frac{4(2\ell+11)(\ell-1)}{2\ell+5}
 \quad\omega_t^2=\omega_0^2\frac{2(2\ell+13)(\ell-1)}{2\ell+7}\quad \omega_0^2=\frac{\mu_c}{\rho_c R^2}\\
  \label{e5.24}
 && \tau_s=\tau_0
 \frac{2(2\,\ell+5)}{(\ell-1)(2\ell+11)}\quad
 \tau_t=\tau_0 \frac{(2\,\ell+7)}{(\ell-1)(2\ell+13)}\quad
 \tau_0=\frac{\rho_c R^2}{\eta_c}.
 \end{eqnarray}
 Note, putting  $\rho_c$ of this non-homogeneous model equal to the density of homogeneous model
 of previous subsection  $\rho$ we find that these stellar models have different
 radius and mass. The considered in this subsection a non-homogeneous model of a solid star is
 bigger in radius and more massive in mass, this difference, as is
 clearly seen, is traced in the values of periods and lifetimes
 for
 both {\it s-mode} and {\it t-mode} of quasistatic oscillations. The presented
 calculation justifies inferences of foregoing subsection
 regarding the frequencies and lifetimes of quasistatic
 spheroidal and torsional non-radial oscillations computed in
 homogeneous and non-homogeneous solid globe models.

\subsection{Self-gravitating mass with non-uniform singular density and non-uniform profiles of shear modulus
and shear viscosity}

 Finally, consider self-gravitating spherical mass of radius $R$ with singular in origin
 density of the form
 \begin{eqnarray}
 \label{e5.25}
 \rho(r)=\frac{5}{6}\rho\left({\frac{R}{r}}\right)^{1/2}.
 \end{eqnarray}
 Unlike the above case of models with regular density profiles, in the last formula the constant
 $\rho$ can be fixed by condition on the globe surface:
 $\rho(r=R)=\rho_s=constant$, not in the center.
 This kind of mass distribution are currently utilized in astrophysical models aimed at the search for dark
 matter candidates (e.g. de Boer, Sander, Zhukov, Gladyshev and Kazakov
 2005). This model is interesting in that the
 total mass
 \begin{eqnarray}
 \label{e5.26}
 {\cal M}=\int \rho(r)\,d{\cal V}= \frac{4\pi}{3}\rho R^3.
 \end{eqnarray}
 has the form identical to that for a spherical mass of a homogeneous
 density. Following the above prescription and computing first the pressure profile inside this object
 from equations of gravitational equilibrium, we obtain
\begin{eqnarray}
 P(r)=P_c\left[1-\left(\frac{r}{R}\right)\right]\quad\quad P_c=\frac{10\pi}{9}G\rho^2\,R^2.
 \label{e5.27}
 \end{eqnarray}
 It is seen that the pressure is linear function of distance from the center and
 in the center the pressure is finite.
 Keeping in mind our conjecture that profiles of shear modulus and shear viscosity follow the same radial dependence as
 the pressure profile does we put
 \begin{eqnarray}
 \mu(r)=\mu_c\left[1-\left(\frac{r}{R}\right)\right]\quad\quad
 \eta(r)=\eta_c\left[1-\left(\frac{r}{R}\right)\right].
 \label{e5.28}
 \end{eqnarray}
 For integral parameters of quasistatic oscillations in spheroidal
 mode we obtain
\begin{eqnarray}
\label{e5.29}
 &&M_s=4\pi\rho\,A_s^2
 R^{2\ell+1}
 \frac{\ell}{2\ell+1}\quad K_s = 4\pi\, A_s^2\,\mu_c\,
 R^{2\ell-1}(\ell-1)\\
 \label{e5.30}
 && D_s = 4\pi
 A_s^2\,\eta_c\, R^{2\ell-1}(\ell-1).
 \end{eqnarray}
The frequency and lifetime are given by
\begin{eqnarray}
\label{e5.31}
 &&
\omega_s^2=\omega_0^2\,(2\ell+1)(\ell-1)/\ell\quad\quad
 \tau_s=\frac{2\tau_0 \ell}{(2\ell+1)(\ell-1)}\\
 && \omega_0^2=\frac{\mu_c}{\rho R^2}\quad \tau_0=\frac{\rho R^2}{\eta_c}.
 \end{eqnarray}
For torsional quasistatic vibrations one has
 \begin{eqnarray}
 \label{e5.32}
 &&  M_t=4\pi\rho\,A_t^2 R^{2\ell+3}\frac{\ell(\ell+1)}{(2\ell+1)(2\ell+3)}\\
 \label{e5.32a}
 &&  K_t =2\pi A_t^2\mu_c\,
 R^{2\ell+1}\frac{\ell(\ell-1)}{(2\ell+1)}
 \quad D_t =2\pi A_t^2\eta_c\,
   R^{2\ell+1}\frac{\ell(\ell-1)}{(2\ell+1)}.
 \end{eqnarray}
For the frequency and lifetime of toroidal mode we get
\begin{eqnarray}
\label{e5.33}
 \omega_t^2=\frac{1}{2}\omega_0^2\frac{(2\ell+3)(\ell-1)}{(\ell+1)},\quad\quad
  \tau_t=\frac{\tau_0(\ell+1)}{(2\ell+3)(\ell-1)}.
 \end{eqnarray}
All the above models preserve one important feature of homogeneous
model regarding the quadrupole ($\ell=2$) degree of lowest
overtone of spheroidal and torsional quasistatic shear
oscillations.

\section{Summary}
 A fundamental understanding of oscillatory modes in a spherical
 mass of viscoelastic solid is needed in different
 areas of physics and astrophysics embracing wide spatial scale of material objects ranging from
 ultrafine pieces of solid matter such as metallic nanoparticles and atomic nuclei to huge astrophysical
 objects like planets of the solar systems and collapsed degenerate compact stars like white dwarfs and
 pulsars. In this paper, the regime of quasistatic shear vibrations
 of viscoelastic solid globe has been investigated in some details whose non-trivial feature is that the
 restoring force of elastic stress and dissipative force of
 viscous stress entering the equation of solid-mechanics turn to zero, whereas the work done
 by the above stresses in the bulk of  oscillating solid globe does not.
 It is shown that in this regime the frequency and lifetime of quasistatic spheroidal and torsional vibrations
 can be computed working from energy balance equation underlying the Rayleigh's energy variational
 method. The efficiency of this method has been demonstrated by analytic results
 for several astrophysical models illustrating the effect of self-gravity
 on profiles of solid-mechanical transport coefficients of shear viscosity and elasticity and its manifestation
 in the spectra of quasistatic non-radial oscillations. The
 practical usefulness of considered mathematical models and obtained analytic estimates
 is that they are quite general and can be applied to different spherical systems whose behavior
 is supposedly governed by equations by solid-mechanics.

\section{Appendices}

 For computational purpose it is convenient to represent
 the components of tensor of shear deformation
 \begin{eqnarray}
 \nonumber
 a_{ik}=\frac{1}{2}(\nabla_i a_k + \nabla_k a_i)
 \end{eqnarray}
 in spherical polar coordinates making use of angle variable ${\zeta}=\hbox{cos}\,{\theta}$.
 The explicit form of these components  reads
 \begin{eqnarray}
 \nonumber
 &&a_{rr}=\frac{\partial a_r}{\partial r}\quad
 \quad\quad
 a_{\theta\theta}=-\frac{(1-\zeta^2)^{1/2}}{r}
 \frac{\partial a_r}{\partial \zeta}+\frac{a_r}{r}\\
 \nonumber
 && a_{\phi\phi}=\frac{1}{r}\frac{1}{(1-\zeta^2)^{1/2}}
 \frac{\partial{a_\phi}}{\partial{\phi}}+\frac{a_r}{r}+
 \frac{\zeta}{(1-\zeta^2)^{1/2}}\frac{a_\theta}{r}
 \\
 \nonumber
 &&a_{r\theta}=\frac{1}{2}\left[-\frac{(1-\zeta^2)^{1/2}}{r}
 \frac{\partial a_r}{\partial \zeta}-\frac{a_\theta}{r}+
 \frac{\partial a_\theta}{\partial r}\right]
 \\
 \nonumber
 &&a_{r\phi}=\frac{1}{2}\left[\frac{1}{r}\frac{1}{(1-\zeta^2)^{1/2}}
 \frac{\partial a_r}{\partial \phi}-\frac{a_\phi}{r}+
 \frac{\partial a_\phi}{\partial r}\right]
 \\
 \nonumber
 &&a_{\theta\phi}=\frac{1}{2}\left[\frac{1}{r}\frac{1}{(1-\zeta^2)^{1/2}}
 \frac{\partial a_\theta}{\partial \phi}-
 \frac{\zeta}{(1-\zeta^2)^{1/2}}\frac{a_\phi}{r}-
 \frac{(1-\zeta^2)^{1/2}}{r}\frac{\partial a_\phi}{\partial \zeta}\right]
 \end{eqnarray}
For the spheroidal mode of quasistatic shear oscillations the
spherical polar components of poloidal field of instantaneous
displacements are give by
 \begin{eqnarray}
 \nonumber
 a_r=A_s\,r^{\ell-1}\ell P_\ell(\zeta)\quad
 a_\theta=-A_sr^{\ell-1}(1-\zeta^2)^{1/2}\frac{dP_\ell(\zeta)}{d\zeta}
 \quad a_\phi=0\qquad \zeta=\cos\theta.
 \end{eqnarray}
The components of the shear deformation tensor in spheroidal mode
of quasistatic oscillations are given by
 \begin{eqnarray}
 \nonumber
&& a_{rr}=A_sr^{\ell-2}\ell(\ell-1)P_\ell(\zeta)\\[0.2cm]
 \nonumber
&& a_{\theta\theta}=A_sr^{\ell-2}
   \left[\zeta\frac{dP_\ell(\zeta)}{d\zeta}-\ell^2P_\ell(\zeta)\right]\\[0.2cm]
 \nonumber
&& a_{\phi\phi}=-A_sr^{\ell-2}
   \left[\zeta\frac{dP_\ell(\zeta)}{d\zeta}-\ell P_\ell(\zeta)\right]\\[0.2cm]
 \nonumber
&& a_{r\theta}=-A_sr^{\ell-2}(\ell-1)
   (1-\zeta^2)^{1/2}\frac{dP_\ell(\zeta)}{d\zeta}\quad a_{r\phi}=0\quad
   a_{\theta\phi}=0
 \end{eqnarray}
After integration over the solid angle for parameters of inertia,
shear stiffness (and shear friction) we obtain
\begin{eqnarray}
 \nonumber
 && M_s=\int\rho(r) a_i\,a_i d{\cal V}=\int \rho(r)
 [a_r^2+a_\theta^2] d{\cal V}= 4\pi\ell A_s^2 \int_0^{R}\rho r^{2\ell}
 \,dr\\ \nonumber
 && K_s = 2 \int \mu(r) a_{ij}a_{ij}\, d{\cal V}
 = 2 \int\mu(r)\left( a_{rr}^2+a_{\theta\theta}^2+a_{\phi\phi}^2+2 a_{r\theta}^2\right) d{\cal V}
 \\ [0.2cm]\nonumber
 && =  8\pi A_s^2\int\limits_{0}^{R}  \mu (r) r^{2\ell-2} dr \int\limits_{-1}^{1}
  \left[
 \ell^2 (\ell^2 -\ell+1) P_\ell(\zeta)^2
 -\ell (\ell+1) \zeta P_\ell(\zeta)\frac{dP_\ell(\zeta)}{d\zeta}+\right. \\ \nonumber
 && \ \  \   \left.
 \zeta^2\left( \frac{d
 P_\ell(\zeta)}{d\zeta}\right) ^{2}
 +(\ell-1) ^2 (1-\zeta^2)\left( \frac{d P_\ell(\zeta)}{d\zeta}\right)^{2}\right]
 d\zeta \\  \nonumber
 &&= 8\pi A_s^2\,\ell(\ell-1)(2\ell-1)\,\int\limits_{0}^{R}
 \mu (r)\, r^{2\ell-2} dr
\end{eqnarray}
and result for $D_s$ is similar to that for $K$. The finale
computational formulas for frequency and lifetime of spheroidal
fundamental mode of quasistatic shear oscillations reads
 \begin{eqnarray}
       \displaystyle \omega_s=\left[\frac{2(2\ell-1)(\ell-1)
 \int\limits_0^{R}\mu(r)r^{2\ell-2}dr}
 {\int\limits_0^{R}\rho(r) r^{2\ell}\,dr}\right]^{1/2}\quad
        \displaystyle \tau_s=\left[\frac{\int\limits_0^{R}\rho(r) r^{2\ell}\,dr}{(2\ell-1)(\ell-1)
  \int\limits_0^{R}\,\eta(r)\,r^{2\ell-2}dr}
  \right].
  \nonumber
 \end{eqnarray}
 In the toroidal fundamental mode of
quasistatic shear oscillations, the  spherical polar components of
toroidal instantaneous displacements have the form
 \begin{eqnarray}
 \nonumber
 a_r=0\quad a_\theta=0\quad a_{\phi}=A_t\,r^\ell(1-\zeta^2)^{1/2}
 \frac{dP_\ell(\zeta)}{d\zeta}
 \end{eqnarray}
The non-zero components of strain tensor in torsional mode are
given by
 \begin{eqnarray}
 \nonumber
&& a_{rr}=a_{\theta\theta}=a_{\phi\phi}=a_{r\theta}=0\\
 \nonumber
 && a_{r\phi}=\frac{1}{2}A_tr^{\ell-1}
   (\ell-1)(1-\zeta^2)^{1/2}\frac{dP_\ell(\zeta)}{d\zeta}\\
 \nonumber
&& a_{\theta\phi}=-\frac{1}{2}A_tr^{\ell-1}
   \left[2\zeta\frac{dP_\ell(\zeta)}{d\zeta}-\ell(\ell+1)P_\ell(\zeta)\right]
 \end{eqnarray}
The inertia, stiffness and friction in torsional mode of
differentially rotational non-radial shear oscillations after
integration over the solid angle have the form
\begin{eqnarray}
 \nonumber
 && M_t=\int\rho(r) a_i\,a_i\,d{\cal V}=\int \rho(r) a_\phi^2\,d{\cal V}=
 4\pi A_t^2 \frac{\ell(\ell+1)}{(2\ell+1)}\int_0^{R}\rho(r) r^{2\ell+2}
 dr\\ \nonumber
 &&K_t = 2\,\int\mu(r)\, a_{ij}a_{ij}\, d{\cal V}
 = 2\,\int\mu(r)\left( a_{r\phi}^2+a_{\theta\phi}^2\right) \,d{\cal
 V}\\ \nonumber
 && = 2\pi A_t^2\int\limits_{0}^{R}  \mu(r)\, r^{2\ell} dr
 \int\limits_{-1}^{1} \left[ \ell^2 (\ell+1)^2 P_\ell(\zeta)^2
 -4\ell (\ell+1)\,\zeta\,P_\ell(\zeta)\frac{dP_\ell(\zeta)}{d\zeta}\right. \\ \nonumber
 &&\ \  \   \left. +\,4\,\zeta^2\left( \frac{d
 P_\ell(\zeta)}{d\zeta}\right) ^{2}
 +(\ell-1) ^2 (1-\zeta^2)\left( \frac{d P_\ell(\zeta)}{d\zeta}\right)^{2}\right]
 d\zeta\\ \nonumber
 &&\ \  \   = 4\pi A_t^2\,\ell(\ell-1)(\ell+1)\,\int\limits_{0}^{R}
 \mu(r)\, r^{2\ell} dr.
 \end{eqnarray}
 The integral coefficient of friction $D$ in torsional mode is computed in a similar fashion.
 Finale computational formula for the frequency and lifetime of toroidal
fundamental mode of quasistatic shear oscillations is given by
  \begin{eqnarray}
 \nonumber
       \displaystyle \omega_t=\left[\frac{(2\ell+1)(\ell-1)
 \int\limits_0^{R}\,\mu(r)\,r^{2\ell}\,dr}
 {\int\limits_0^{R}\,\rho(r)\,r^{2\ell+2}\,dr}\right]^{1/2}\quad
 \displaystyle \tau_t=\left[\frac{2\int\limits_0^{R}\,\rho(r)\,r^{2\ell+2}\,dr}
 {(2\ell+1)(\ell-1)\int\limits_0^{R}\,\eta(r)\,r^{2\ell}\,dr}\right].
  \end{eqnarray}
 The practical usefulness of the calculus presented in this appendices
 is that it can be utilized in the study of diverse class of solid globe models of physical
 interest.

\section*{Acknowledgment}
This work is partially supported by NSC of Taiwan, Republic of
China, under grants NSC 96-2811-M-007-001 and NSC-95-211-M-007-050
and under protocol JINR (Russia)-- IFIN-HH (Romania), project
number 3753-2007.


\begin{thebibliography}{00}


\bibitem{B-99} H. Bethe, Nuclear Physics, {\it Rev. Mod. Phys.} {\bf 71} (1999) S6--S15.



\bibitem{Guyer-FS-73} R.A. Guyer, The Fermi solid, in {\it Selected Topics in Physics, Astrophysics, and
Biophysics} (Reidel, Dordrecht-Holland, 1973) pp. 44-49.

\bibitem{QS-03} E.Polturak and N.Gov, Inside a quantum solid, {\it Contemporary Physics} {\bf 44} (2003) 145--151.



\bibitem{MH-06} M. Harwit,  {\it Astrophysical Concepts} 3th edn (Springer-Verlag, 2006).

\bibitem{ST-83} S.L. Shapiro and S.A. Teukolsky, {\it Black Holes, White Dwarfs and Neutron Stars}
                  (Wiley, New York, 1983).


\bibitem{FW-99} F. Weber, {\it Pulsars as astrophysical laboratory for nuclear and particle physics}
                          (IOP Publishing 1999).



\bibitem{G-99} N. Glendenning, {\it Compact Stars} (Springer, Berlin, 2000).


\bibitem{LP-99} J.M. Lattimer and M. Prakash, Neutron Star Structure and the Equation of State, {\it Astrophys. J.}
                {\bf 550} (2001) 426--442.



\bibitem{W-05} F. Weber, Strange quark matter and compact stars,
               {\it Prog. Part. Nucl. Phys.} {\bf 54} (2005)
               193---288.

\bibitem{XU-03} R.X. Xu, Solid quark stars,
               {\it Astrophys. J.} {\bf 596} (2003) L59--L62.

\bibitem{O-05} B.J. Owen, Maximum elastic deformations of compact stars with exotic equations of etate
               {\it Phys. Rev. Lett.} {\bf 95} (2005) 211101.


\bibitem{VH-80} H.M. van Horn, Micropulses, drifting subpulses, and nonradial oscillations of neutron stars,
                {\it Astrophys. J.} {\bf 236} (1980) 899--903.



\bibitem{MVH-88} P.N. McDermott, H.M. van Horn and C.J. Hansen,
 Non-radial oscillations of neutron stars, {\it Astrophys. J.},
 {\bf 325} (1988) 725--748.

\bibitem{SVHC-92} T.E. Strohmayer, J.M. Cordes and H.M. van Horn,
 Determining the coherence of micropulses, {\it Astrophys. J.},
 {\bf 389} (1992) 685--694.

\bibitem{BWP-99} S.I. Bastrukov, F. Weber and D.V. Podgainy, On
  stability of global non-radial pulsations of neutron stars {\it J. Phys.} {\bf G25} (1999) 107--127.


\bibitem{PSR-74} D. Pines, J. Shaham and M.A. Ruderman, Neutron Star Structure from Pulsar
Observations,
 in {\it Physics of Dense Matter} Proc. of IAU Symposium No. 53 (1974) 189--207.

\bibitem{CanChi74} V. Canuto and S.M. Chitre, Crystallization of dense neutron matter,
                   {\it Phys. Rev.} {\bf D9} (1974) 1587--1613 .


\bibitem{BBGM-89} O. Blaes, R. Blandford, P. Goldreich, P. Madau, Neutron starquake models for gamma-ray bursts,
 {\it Astrophys. J.} {\bf 343} (1989) 839--848.

\bibitem{LLE-00} L.M. Franco, B. Link and R.I. Epstein, Quaking neutron stars, {\it Astrophys. J.}
               {\bf 543} (2000) 987--994.



\bibitem{EPS-96} B. Cheng, R.I. Epstein, R.A. Guyer and  A. C.Young,
                 Earthquake-like behaviour of soft gamma-ray
                 repeaters,
                 {\it Nature} {\bf 382} (1996) 518--520.


\bibitem{R-68} M. Ruderman, Crystallization and torsional oscillations of superdense
stars, {\it Nature} {\bf 218} (1968) 1128--1129.


\bibitem{HC-80} C.J. Hansen and D.F. Cioffi, Torsional oscillations in neutron star crusts, {\it Astophys. J.}
               {\bf 238} (1980) 740--742.

\bibitem{BPYW-02} S. Bastrukov, D. Podgainy, J. Yang and F. Weber,
                  Magneto Torsional Pulsations of Magnetars, {\it Mem. Soc. Astron. Ital.} {\bf 73} (2002) p. 522--528.



\bibitem{HKC-05} Lupin C.-C. Lin and Hsiang-Kuang Chang, Periodicity Search in the X-ray Data of RX J0007.0+7302,
                  {\it Astrophys. and Space Science} {\bf 267} (2005)
                 361--367.


\bibitem{SW-06} T.E. Strohmayer and A.L. Watts, The 2004 hyperflare from SGR 1806-20: Further evidence for global
              torsional vibrations, {\it Astrophys. J.}, {\bf 653} (2006) 593--601.

\bibitem{SB-06} T. Strohmayer and L. Bildsten, New views of thermonuclear
 bursts, in {\it Compact stellar X-ray sources} Cambridge Astrophys. Series No. 39
 (2006)  113--156.



\bibitem{GBK-06} G.S. Bisnovatyi-Kogan, Dynamic confinement of jets by magnetotorsional
oscillations,
 {\it Monthly Notes Roy. Astronom. Soc.}, {\bf 376} (2007) 457--464.




\bibitem{B-74} G.F. Bertsch,  Elasticity in the response of nuclei, {\it Ann. Phys.
 N.Y.} {\bf 86} (1974) 138--146.




\bibitem{HD-03} M.N. Harakeh and A. van der Woude, {\it Giant Resonances: Fundamental High-Frequency Modes of Nuclear
                 Excitations} (Oxford Science Publishing, Oxford, 2003).


\bibitem{R-03} A. Richter, Giant resonances - experiments at highest resolution, wavelets and
               scales, {\it Prog. Part. Nucl. Phys.} {\bf 55} (2005) 387--396.


\bibitem{S-77} S.F. Semenko, Collective coordinates in the random phase approximation and the systematics of giant
               multipole resonances, {\it Sov. J. Nucl. Phys.} {\bf 26} (1977) 143--146.






\bibitem{SH-78} H. Sagawa and G.  Holzwarth,
               Giant Resonances: A Comparison between TDHF and Fluid Dynamics in Small Amplitude Vibrations
               of Spherical Nuclei, {\it Prog. Theor. Phys.} {\bf A325} (1978) 1213--1229.




\bibitem{NS-80} J.R. Nix and A.J.  Sierk,
 Macroscopic description of isoscalar giant multipole resonances,
 {\it Phys. Rev.} {\bf C21} (1980)  396--404.


 \bibitem{WA-81} C.-Y. Wong and  N. Aziz, Nuclear giant resonances as elastic vibrations, {\it Phys.
 Rev.} {\bf C24} (1981) 2290--2310.


 \bibitem{HGWL-82} R. Hasse,  G. Ghosh, J. Winter and A. Lumbroso,
 Isoscalar giant-resonance energies and long-mean-free path nuclear fluid
 dynamics, {\it Phys. Rev.} {\bf C25} (1981) 2771--2779.

\bibitem{S-83} S. Stringari, Fluid-dynamical description of nuclear collective excitations, {\it Ann. Phys.} (N.Y.)
               {\bf 151} (1983) 35--70.

\bibitem{N-84} W. Norenberg, Elastoplasticity  of finite fermi systems,
               {\it Nucl. Phys.} {\bf 428} (1984) 177--187.


\bibitem{BM-88} E.B. Balbutsev and I.N. Mikhailov,
 Dynamics of nuclear integral characteristics, {\it J. Phys.} {\bf
 G14} (1988) 545--567.

\bibitem{KOL-90} V.M. Kolomietz, {\it Local density approach to atomic and nuclear physics}
 (Naukova Dumka, Kiev, 1990).

\bibitem{DT-91} M. Di Toro, Semiclassical approach to the description of collective nuclear motions,
 {\it Sov. J. Part. Nucl.} {\bf 22} (1991) 185--210.

\bibitem{BMS-93} S.I. Bastrukov, \c S. Mi\c sicu and A.V. Sushkov,
 The dipole torus mode in nuclear fluid-dynamics, {\it Nucl. Phys.} {\bf A562} (1993) 191--204.


\bibitem{SM-06} \c S. Mi\c sicu,
 Interplay of compressional and vortical nuclear currents in overtones of the isoscalar giant dipole
 resonance, {\it Phys. Rev.} {\bf C73} (2006) 024301.

\bibitem{RFH-91} S. Raman, L.W. Fagg and R. Hicks, Giant Magnetic Resonances
 in: {\it Electric and Magnetic Giant Resonances} ed. by J. Speth
 (World Scientific, Singapore, 1991) p.355.

\bibitem{LNCNR-96} C. Luttge, P. von Neumann-Cosel, F. Neumeyer and A. Richter,
 Magnetic dipole and quadrupole response of nuclei, supernova physics and in-medium vector meson
 scaling,
 {\it Nucl. Phys.} {\bf A606} (1996) 183--200.


\bibitem{HE-77} G. Holzwarth and G. Eckart, The nuclear "Twist",
                {\it Z. Phys.} {\bf A283} (1977) 219--220.


\bibitem{BG-91} S.I. Bastrukov and V.V. Gudkov, The twist $M\lambda,T=0$ giant modes in spherical
                nuclei, {\it Z. Phys.} {\bf A341} (1992) 395--399.

\bibitem{BMS-95} S.I., Bastrukov, I.V. Molodtsova and V.M. Shilov,
 Torsional multipole magnetic response of a heavy spherical
 nucleus,
 {\it Phys. Scripta} {\bf 51} (1995) 54--59.


 \bibitem{BLM-97} S.I. Bastrukov, J. Libert and  I. Molodtsova,
 Elastodynamic features of nuclear matter from macroscopic model of giant magnetic
 resonances, {\it Int. J. Mod. Phys.} {\bf E6} (1997) 89--110.


\bibitem{GM-96} W. Greiner and J.A. Maruhn, {\it Nuclear Models} (Springer, Berlin,
 1996).

\bibitem{ROWE-70} D.J. Rowe, {\it   Nuclear collective motion: models and theory} (Methuen, London, 1970).


\bibitem{LAMB} H. Lamb, On the vibrations of an elastic sphere, {\it Proc. London. Math. Soc.} {\bf 13}
   (1881-82) 189--212.


\bibitem{LOVE} A.E.H. Love,  {\it Treatise on Mathematical theory of elasticity} (Dover, New York,
 1944).


\bibitem{MF-56} R.M. Morse and H. Feshbach, {\it Methods of Mathematical Physics} (MCGraw-Hill, New York, 1953).


\bibitem{ES-EL} A.C. Eringen and E.S. Suhubi, {\it Elastodynamics} vol. 2 (Pergamon Press, New York, 1975).


\bibitem{LL-EL} L.D. Landau, E.M. Lifshits, A.M. Kosevich and L.P. Pitaevskii, {\it Theory of
                  Elasticity} (Butterworth-Heinenann, Oxford 1995).


\bibitem{CHANDRA} S. Chandrasekhar, {\it Hydrodynamic and Hydromagnetic
  stability} (Clarendon Press, Oxford, 1961).




\bibitem{LAMB-H} H. Lamb,  {\it Hydrodynamics} 6th edn (Dover, New York, 1945).


\bibitem{Jef-70} H. Jeffreys, {\it The Earth}, 6th edn (Cambridge University Press, 1976).


\bibitem{LU-81} R.R. Lapwood and  T. Usami, {\it Free Oscillations
               of the Earth} (Cambridge University Press, 1981).



\bibitem{AR-02} K. Aki and  P.G. Richards, {\it Quantitative Seismology} (University Science Books,
Sausalito, California, 2002).


\bibitem{H-03} J. Hern\'andez-Roses, M. Picquart, E.
Haro-Poniatowski, M. Kanehisa, M. Jouanne and J.F.
 Morhange, Elastic vibrations of spheroidal nanometric particles, {\it J. Phys.: Cond. Matter} {\bf 15} (2003)
 7481--7487.

\bibitem{DMurray-2005} E. Duval, L. Saviot, A. Mermet and  D.B. Murray,
Continuum elastic sphere vibrations as a model for low lying
optical modes in icosahedral quasicrystals, {\it J. Phys.: Cond.
Matter} {\bf 17} (2005) 3559--3565.


\bibitem{BL-05} S.I. Bastrukov and Pik-Yin Lai,
   Resonant response of a metallic nanoparticle by collective
   cyclotron oscillations of electrons and ions, {\it Phys. Lett.}
   {\bf A341} (2005) 207--211.


\bibitem{BAST-94} S.I. Bastrukov, Low frequency elastic response of a spherical particle, {\it Phys. Rev.} {\bf E49}
    (1994) 3166-3170.



\bibitem{Ye-Zhen} Zhen Ye, On the low frequency elastic response of a spherical
         particle, {\it Chinese J. Phys.} {\bf 38} (2000) 103--110.






\bibitem{H-78} R. Hasse, Approaches to nuclear friction, {\it Rep. Prog. Phys.}
                 {\bf 41} (1978) 1027--1101.


 \bibitem{FW}  A.L. Fetter and J.D. Walecka,
               {\it Theoretical Mechanics of Particles and
               Continua} (McGraw-Hill,  New York, 1980).



\bibitem{BAST-AST97} K. Alder and R. M. Steffen, Emission and
  Absorption of electromagnetic radiation. in {\it
  The Electromagnetic Interaction in Nuclear Spectroscopy}
  (North-Holland, Amsterdam, 1975) 1--55.




\bibitem{FI-76} E. Flowers and N. Itoh, Transport properties of dense matter, {\it
 Astrophys. J.} {\bf 206} (1976)  218--242.



\bibitem{CL-87} C. Cutler and L. Lindblom,
 The effect of viscosity on neutron star oscillations, {\it
 Astrophys. J.} {\bf 314} (1987) 234--241.



\bibitem{JOP-03} S.I. Bastrukov, J. Yang, D.V. Podgainy and F.
Weber, Signatures of field induced spin polarization of neutron
star matter in seismic vibrations of paramagnetic neutron star,
{\it J. Phys.} {\bf G29} (2003) 683--695.








\bibitem{B-93} S.I. Bastrukov, On elastic vibrations of homogeneous
star, {\it Mod. Phys. Lett.} {\bf A8} (1993) 711--714.



\bibitem{BAST-96} S.I. Bastrukov, Non-radial vibrations of a star modeled
 by a heavy spherical mass of an elastic solid, {\it Phys. Rev.} (1996) {\bf E53}
 1917--1922.



\bibitem{DM-05} W. de Boer, C. Sander, V. Zhukov, A. Gladyshev and D. Kazakov,
EGRET excess of diffuse galactic gamma rays as tracer of dark
matter, {\it Astron. and Astrophys.} {\bf 444} (2005) 51--67.

 \end{thebibliography}
\end{document}